\documentclass[sigconf]{acmart}

\graphicspath{{Figures/}}

\acmConference[CHI '24]{Proceedings of the CHI Conference on Human Factors in Computing Systems}{May 11--16, 2024}{Honolulu, HI, USA}
\acmBooktitle{Proceedings of the CHI Conference on Human Factors in Computing Systems (CHI '24), May 11--16, 2024, Honolulu, HI, USA}
\copyrightyear{2024}
\acmYear{2024}
\setcopyright{rightsretained}
\acmDOI{10.1145/3613904.3642651}
\acmISBN{979-8-4007-0330-0/24/05}

\usepackage{color}
\usepackage{multirow}
\usepackage{amsmath}
\usepackage{xcolor}

\newcommand{\ana}[1]{\textcolor{black}{#1}}
\newcommand{\laia}[1]{\textcolor{black}{#1}}
\newcommand{\amar}[1]{\textcolor{black}{#1}}
\newcommand{\marte}[1]{\textcolor{black}{#1}}
\definecolor{cadet}{rgb}{0.33, 0.41, 0.47}

\begin{document}

\title[SoniWeight Shoes: Effects and Personalization of a Wearable Sound Device for Altering Body Perception and Behavior]{SoniWeight Shoes: Investigating Effects and Personalization of a Wearable Sound Device for Altering Body Perception and Behavior}

\author{Amar D'Adamo}
\orcid{0009-0005-5863-692X}
\affiliation{
 \institution{Universidad Carlos III de Madrid}
  \city{Madrid}
   \country{Spain}}
\email{adadamo@inf.uc3m.es}

\author{Marte Roel Lesur}
\orcid{0000-0003-2935-4868}
\affiliation{
 \institution{Universidad Carlos III de Madrid}
  \city{Madrid}
   \country{Spain}}
\email{mroel@inf.uc3m.es}

\author{Laia Turmo Vidal}
\orcid{0000-0002-1769-0138}
\affiliation{
 \institution{Universidad Carlos III de Madrid}
  \city{Madrid}
   \country{Spain}}
\email{laia.turmo@uc3m.es}

\author{Mohammad Mahdi Dehshibi}
\orcid{0000-0001-8112-5419}
\affiliation{
 \institution{Universidad Carlos III de Madrid}
  \city{Madrid}
   \country{Spain}}
\email{dmohamma@inf.uc3m.es}

\author{Joaquín Roberto Díaz Durán}
\orcid{0000-0002-7180-0721}
\affiliation{
 \institution{Universidad Carlos III de Madrid}
  \city{Madrid}
   \country{Spain}}
\email{jodiazd@pa.uc3m.es}

\author{Daniel De La Prida}
\orcid{0000-0002-4386-5544}
\affiliation{
 \institution{Universidad Carlos III de Madrid}
  \city{Madrid}
   \country{Spain}}
\email{dprida@ing.uc3m.es}

\author{Luis Antonio Azpicueta-Ruiz}
\orcid{0000-0002-2458-0914}
\affiliation{
 \institution{Universidad Carlos III de Madrid}
  \city{Madrid}
   \country{Spain}}
\email{lazpicue@ing.uc3m.es}

\author{Aleksander Väljamäe}
\orcid{0000-0001-6071-3211}
\affiliation{
 \institution{University of Tartu}
  \city{Tartu}
   \country{Estonia}}
\email{aleksander.valjamae@gmail.com}

\author{Ana Tajadura-Jiménez}
\orcid{0000-0003-3166-3512}
\affiliation{
 \institution{Universidad Carlos III de Madrid}
  \city{Madrid}
   \country{Spain}}
\email{atajadur@inf.uc3m.es}

\renewcommand{\shortauthors}{D'Adamo et al.}

\begin{abstract}

Changes in body perception influence behavior and emotion and can be induced through multisensory feedback. Auditory feedback to one's actions can trigger such alterations; however, it is unclear which individual factors modulate these effects. We employ and evaluate SoniWeight Shoes, a wearable device based on literature for altering one's weight perception through manipulated footstep sounds. In a healthy population sample across a spectrum of individuals (n=84) with varying degrees of eating disorder symptomatology, physical activity levels, body concerns, and mental imagery capacities, we explore the effects of three sound conditions (low-frequency, high-frequency and control) on extensive body perception measures (demographic, behavioral, physiological, psychological, and subjective). Analyses revealed an impact of individual differences in each of these dimensions. Besides replicating previous findings, we reveal and highlight the role of individual differences in body perception, offering avenues for personalized sonification strategies. Datasets, technical refinements, and novel body map quantification tools are provided.
\end{abstract}




\begin{teaserfigure}
\begin{center}
  \includegraphics[width=15.3cm]{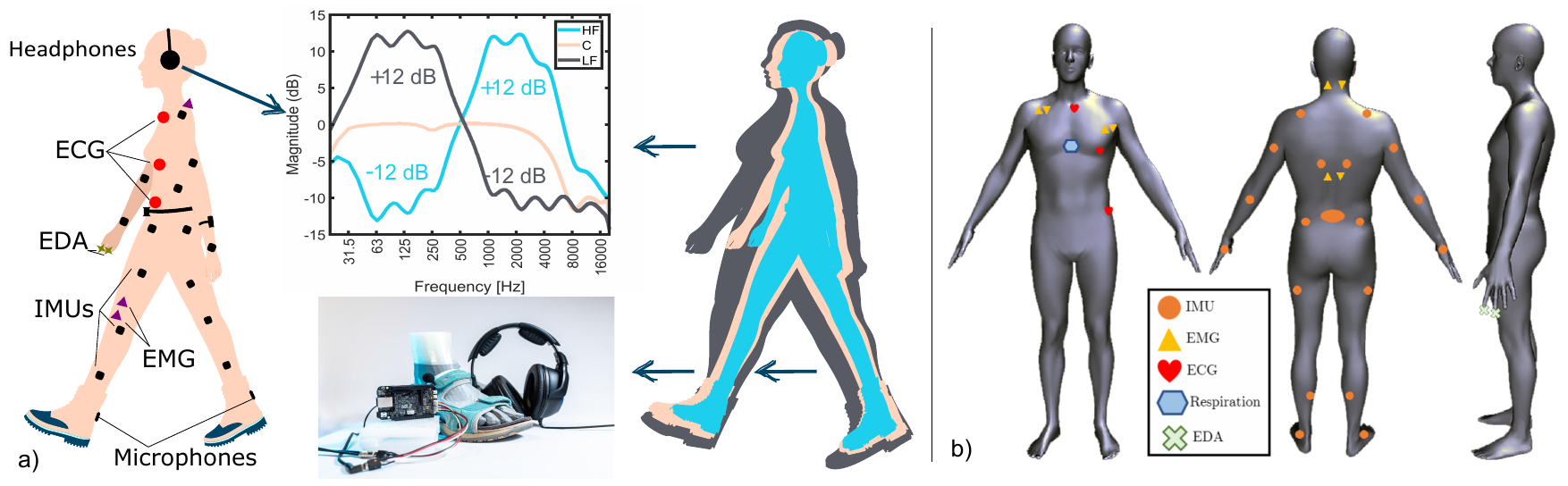}
  \caption{a): Overview of the sound device to induce an illusion of changes in body weight, including the detail of the placement of sensors used to track gait and physiological responses (IMU=inertial measurement unit, EMG=electromyogram, EDA=electrodermal activity, ECG=electrocardiogram), and the filters used in the three sound conditions: HF (High frequency), LF (Low frequency) and C (control). b): Body views with the locations of the sensors used during the experiment.}
  \label{fig:teaser}
  \Description{The picture shows an overview of the operation of SoniWeigth device: The participant is wearing EMG, ECG, EDA sensors and a full body motion tracksuit. The participant is also wearing microphones on the shoes and headphones, trough the latter, modified sounds of the footsteps are reproduced, according to three different audio filters. The sounds produce sensations of being in a lighter/heavier body.}
  \end{center}
\end{teaserfigure}


\maketitle

\section{Introduction}
\laia{The appearance and development of body sensing and actuating technologies has prompted research on the potential of multisensory technologies to alter body perception \cite{tajadura-jimenez_action_2012}.} Body perception is a multifaceted phenomenon \marte{encompassing} both perceptual (e.g., perceived body size) and attitudinal (e.g., emotional \ana{responses} and thoughts towards body size) components~\cite{Dijkerman2018} \marte{that impact} motor \ana{\cite{cardinali_tool-use_2009}}, emotional\ana{~\cite{Pollatos2008, Schwarzer1992}}, and social functioning\amar{~\cite{tacikowski_fluidity_2020, clausen_action_2021, Bedder2019}.} For example, negative body perceptions and appraisals \marte{regarding one's} body size, shape, appearance or capabilities are intricate with many health concerns~\cite{gallagher_how_2005, Schwarzer1992, Pollatos2008, andrew_positive_2016}, such as eating disorders~\cite{Cash1997} and physical inactivity~\cite{Schwarzer1992, Cairney2007, McAuley1993}. Multisensory stimulation related to \marte{one's} body \laia{can drastically alter perception }\marte{\cite{botvinick_rubber_1998, LESUR2023108477,tajadura-jimenez_action_2012}}, which in turn affects emotions, cognition and behavior~\cite{Dijkerman2018, tajadura-jimenez_as_2015, clausen_action_2021}. 

An emerging approach within HCI utilizes auditory feedback to alter body perception~\cite{tajadura-jimenez_principles_2018,Spence2020}, which has the benefit of being transparent, portable, ubiquitous and suitable for in-the-wild contexts. Using metaphorical sounds like wind~\cite{ley-flores_soniband_2021, ley-flores_altering_2019}, musical notes~\cite{Singh:2014:MPC:2611247.2557268,Singh2016,Newbold2016}, pitch variations~\cite{ley-flores_effects_2022, Nava2020,Tajadura2017contingent} or modifying natural action sounds such as tapping~\cite{tajadura-jimenez_action_2012,tajadura_action_2015}, or walking sounds~\cite{tajadura_action_2015, tajadura-jimenez_as_2019}, has shown the potential to influence body perception and related attitudes. \marte{However, previous works have often relied on narrow population groups (e.g.,~\cite{tajadura-jimenez_body_2022, tajadura-jimenez_bodily_2017,gomez-andres_enriching_2020}) or small sample sizes (e.g.,~\cite{tajadura-jimenez_bodily_2017, ley-flores_soniband_2021,Singh:2014:MPC:2611247.2557268}), which impedes assessing how potentially relevant target groups might differently react to body sonification.}

We set to formally test the potential of sound feedback to alter body perception \marte{in a healthy population sample across a spectrum of individuals. We applied} a comprehensive experimental study of SoniWeight Shoes, a novel device based on~\cite{tajadura-jimenez_as_2015} that alters the frequency of footstep sounds to induce body-weight illusions. Related work has shown that manipulating footstep sounds during walking induces changes in perceived body weight in the general population~\cite{tajadura-jimenez_as_2015,tajadura-jimenez_as_2019}, and people with eating disorders~\cite{tajadura-jimenez_body_2022}. To assess the influence of individual factors on changes in body perception induced by sound, this study involved n=84 participants stratified into four equally-sized groups based on individual levels of symptomatology of eating disorders (SED) and physical activity (PA) levels. \marte{We aimed to investigate how SED, PA levels, body concerns (encompassing both body appearance and body capabilities), gender perceptions, and individual levels of sensory imagery (i.e., the capacity to imagine situations across sensory modalities) may modulate the effects of sounds on body perception.} Participants engaged in a walking task while listening to their own footsteps in three conditions differing by the applied filtering: a control condition with no filtering, a filter enhancing high frequencies (\marte{i.e.,} sounds consistent with a lighter body), and one enhancing low frequencies (\marte{i.e.,} sounds consistent with a heavier body). To investigate how each of these sonification conditions contributes to body perception, we included a variety of measures: demographic and personal (e.g., levels of PA, SED, and body concerns), behavioral (e.g., gait, body visualization), physiological (e.g., electrodermal activity, muscular activity), psychological (body concerns, sensory imagery) and subjective reports (e.g., questionnaires and body maps~\cite{turmo_vidal_towards_2023}). 

This work presents three main contributions: 
\begin{itemize}
    \item {Insights into how SoniWeight Shoes-induced changes in body perception are modulated by individual differences. This, for SED, PA, body concerns, gender perception and sensory imagery, across \marte{the mentioned set of measures. This exploration gains significance against the backdrop of pervasive negative body perceptions linked to both SED~\cite{Cash1997} and physical inactivity~\cite{Schwarzer1992, Cairney2007, McAuley1993}, prevalent societal problems}}.  
     \item Replication of previous work \ana{on sonification strategies for altering body perception} across various measures and with \marte{a more diverse and sizable cohort}, both for the general population~\cite{tajadura-jimenez_as_2015} and people with a relatively high degree of SED~\cite{tajadura-jimenez_body_2022}. This is an important contribution given the lack of reproducibility and inconsistency across measures often present in this field (e.g.,~\cite{DEHAAN201768, Mottelson2023}).
    \item An extensive dataset, made accessible to the research community, incorporating demographic, personal and psychological data, along with physiological, behavioral and subjective experimental data. This dataset supports the study of particular individual factors and provides an essential foundation for future work. 
\end{itemize}
 \laia{This study contributes empirical knowledge to research on sonification for body perception (e.g. ~\cite{tajadura-jimenez_as_2015,tajadura-jimenez_as_2019,brianza_as_2019,clausen_action_2021, gomez-andres_enriching_2020}) by formally addressing the role of individual differences.}
Two additional contributions are: \marte{the user evaluation of the SoniWeight Shoes' prototype~\cite{tecniacustica_delaprida2022} that, with its technical improvements, now is digital, more portable and lighter than~\cite{tajadura-jimenez_as_2015}}; and a methodological novelty in the use of quantitative body maps~\cite{turmo_vidal_towards_2023} as a tool for HCI research. We offer insights for HCI to support sonification strategies, and to \laia{ fundamental research on the link between different traits and changes in body perception. We conclude with a call for personalization of sensory feedback, highlighting the importance for the regular inclusion of individual differences in related research.}

\section{Background/Related Work}

\subsection{Body Perception and Bodily Illusions}
Body perception results from the integration of multisensory signals (such as those emerging from sight, hearing, touch and beyond) and prior knowledge regarding one's body.~\cite{blanke2012, blanke2015, Apps2014}. This capacity allows us to keep track and adjust to the arrangement and position of our body parts and our continuously changing appearance and dimensions~\cite{de_vignemont_mind_2018}. Beyond our kinesthetic functioning, the way we perceive our body is related to cognitive (e.g., prior beliefs; social perception;~\cite{Apps2014,Bedder2019, DURLIK201442}, affective or emotional processes (e.g., self-concept and self-esteem,~\cite{Pollatos2008, tacikowski_fluidity_2020, tajadura-jimenez_balancing_2014, clausen_action_2021}. For instance, when we feel happy, we tend to perceive our bodies as lighter and more agile; conversely, we may perceive our bodies as heavier and more sluggish when we feel sad~\cite{Hartmann2023}. 

Neuroscientific studies have repeatedly shown that people may perceive dummy limbs~\cite{botvinick_rubber_1998}, tools~\cite{maravita2004}, or virtual bodies~\cite{kilteni_extending_2012,tajadura-jimenez_embodiment_2017} as constitutive of their own bodies within seconds of multisensory stimulation. A widely used experimental paradigm is the Rubber Hand Illusion~\cite{botvinick_rubber_1998}, in which participants feel ownership of a rubber hand that is stroked in time with their own hand, which is hidden from view. Here, the integration of proprioceptive, tactile, and visual body signals leads to the illusion of a rubber hand \marte{belonging to one's body~\cite{Tsakiris2005,Ehrsson2020}}. New sensor-based and bodily sensory feedback devices, such as those brought by immersive virtual reality, open new possibilities for creating such Body Transformation Experiences. For instance, creating the illusion of having a longer arm~\cite{kilteni_extending_2012}, a shorter/taller slimmer/wider body or embodying a child’s body~\cite{van_der_hoort_being_2011, tajadura-jimenez_embodiment_2017, Piryankova_weight}.  

\subsection{Changing Body Perception Through Sound}
Auditory feedback contributes fundamentally to the perception of one's body and its surrounding space~\cite{kitawaga2006}. This is highlighted for self-produced sounds or sounds deriving from actions~\cite{Aglioti2010}. Acoustic feedback has been used to change the perception of one’s limbs, including their length (e.g., arms~\cite{tajadura-jimenez_action_2012, tajadura_action_2015}, legs~\cite{tajadura-jimenez_principles_2018}, fingers~\cite{tajadura-jimenez_bodily_2017, Rick2022}, materiality (e.g., the marble hand illusion~\cite{Senna2014}, to alter full body weight perception~\cite{tajadura-jimenez_as_2015} and to induce different feelings about the body or its movement~\cite{ley-flores_altering_2019, ley-flores_soniband_2021, ley-flores_effects_2022, Singh:2014:MPC:2611247.2557268, Singh2016, Newbold2016}. Critically, these studies have also shown that people adapt their behavior to their newly perceived body, as when performing reaching movements~\cite{tajadura2016arm,tajadura_ball2018}.

\subsubsection{Altering Perceived Body Weight through Sound: the Footsteps Illusion}
Most relevant to our work, is the work by~\cite{tajadura-jimenez_as_2015}, who designed a technological prototype to change the frequency spectra of participants’ footstep sounds in real-time. Short-term use of this device while walking on a flat surface was found to alter participant’s perceived body weight. Shifting the sound of the footsteps to lower frequencies resulted in the perception of a heavier/wider body, while shifting the sound to higher frequencies resulted in the perception of a lighter/slimmer body. The induced alterations in body perception were further linked to \ana{emotional responses} and changes walking behavior: higher frequency footsteps were linked to increased happiness, and more dynamic walking patterns with shorter heel strikes. The same device has been used with clinical populations achieving promising results~\cite{tajadura-jimenez_bodily_2017, gomez-andres_enriching_2020}. \marte{These findings are in line with alternative manipulations of footstep sounds that have shown to influence walking behavior~\cite{Bresin2010, menzer_feeling_2010, Turchet2013}}.

Further investigations revealed that the effects of the acoustic manipulation varied based on other factors. Specifically, individuals with high SED levels exhibited contrasting outcomes compared to those with low or medium symptomatology~\cite{tajadura-jimenez_body_2022}. Additionally, the footsteps illusion's impact was influenced by participants' body weight and masculinity/femininity aspirations, but not the participants' gender~\cite{tajadura-jimenez_as_2019}. Although studies on the effects of this illusion with respect to individual levels of physical activity are lacking, we hypothesize such interactions may exist based on other study findings on the effects of movement sonification on body perception~\cite{ley-flores_soniband_2021}. We further \marte{hypothesize that the effects of sounds on body perception may also be modulated by differences in body concerns}. Below, we expand on the theoretical grounds \marte{for each hypothesis} and the relevance of personalizing the technology.

\subsection{Personalization of Technology to Foster Changes in Body Perception, Behavior and Emotion}
Anomalous bodily experiences and integration of bodily signals have been reported in many clinical conditions, such as in eating disorders. \marte{Additional individual factors may influence how sensory cues influence} body perception~\cite{Navas2023}. However, despite the vast literature in the field, relatively \marte{few works have systematically studied this influence}. At large, the novel contribution of this work is to study how the malleability of body perception may be influenced by individual differences in SED, PA and body concerns. This is an important step in personalization and translating this line of work into clinically relevant applications that are sensitive to participants' own needs. 

\subsubsection{Eating Disorders}
Distorted body perceptions have been reported in people with \marte{clinical levels of SED. A study showed that people with} Anorexia Nervosa rotated their shoulders \marte{more widely than their shoulder width} to walk through door openings, \marte{suggesting a disturbed perception of body size influencing} motor behavior~\cite{keizer_too_2013}. \marte{Other studies} have reported an increased plasticity of body perception during visuo-tactile bodily illusions~\cite{Eshkevari2012, Keizer2014}\marte{, pointing at an alteration in the processing of bodily signals for Anorexia Nervosa compared to controls}. However, it is not clear whether \marte{this apparently increased malleability of body perception is caused by an overreliance on visual bodily signals, or by a more general alteration of multisensory integration in the context of body perception.} 
Using the footsteps illusion,~\cite{tajadura-jimenez_body_2022} showed that participants with Anorexia Nervosa and SED displayed a gait typical of heavier bodies and a widest/heaviest visualized body in the sound condition signaling a ‘light’ body, contrasting with healthy participants who showed these patterns in the ‘heavy’ footstep condition. These results suggest disturbances in the sensory integration mechanisms in people with SED, but further investigations with a larger sample of participants are needed and therefore put forth in this study.

\subsubsection{Physical Activity}
Negative body perceptions are a strong predictor of physical inactivity (PA), which itself is a critical health risk~\cite{Jankauskiene2019}. Insufficient physical exercise ranks as the fourth greatest cause for mortality worldwide~\cite{WHO}. Prior works have shown that activity-tracking technologies, rooted in personal informatics~\cite{Kersten-vanDijk2017,Rooksby2014} and behavioral change techniques~\cite{Harrison:2014:TPA:2638728.2641320}, may not alone effectively promote PA~\cite{Kersten-vanDijk2017}, but incorporating psychological factors (e.g., self-esteem, body perception, motivation, emotions) is critical to tackle the problem~\cite{Biddle2007,Rick2022}. Recent studies have explored sensory feedback, including the footsteps illusion~\cite{tajadura-jimenez_as_2015} and movement sonification~\cite{ley-flores_altering_2019,ley-flores_soniband_2021}, as a means to enhance PA by altering body perception. A study relying on an alternative body sonification strategy highlighted differential effects of sound on physically active and inactive people~\cite{ley-flores_soniband_2021}. Yet, it had a limited sample size (5 inactive vs. 7 active participants) and used only qualitative methods. To bridge this gap, our study employs quantitative approaches and a comprehensive dataset to examine the effects of sound across both inactive and active individuals.

\subsubsection{Body concerns}
The pervasive issue of negative body perceptions poses a significant societal problem. Negative body perceptions are entangled with many health conditions, including eating disorders~\cite{Cash1997} and physical inactivity~\cite{Schwarzer1992, Cairney2007, McAuley1993}, and these are known to influence their motor, social, or emotional functions~\cite{Schwarzer1992,Tinetti1993, Pollatos2008, Treasure2010}. As a result, many clinical interventions recommend targeting negative body perceptions~\cite{Tinetti1993}. As mentioned before, prior beliefs about one's body influence body perception~\cite{blanke2012, blanke2015, Apps2014}. Given the link to negative body concerns and mental health conditions, we aimed to investigate how such concerns might modulate the illusion. We expected that the acoustic manipulation would have a lesser effect on participants with a high degree of body concerns due to their relative weighting on prior beliefs and expectations rather than on ongoing sensory signals. 
\subsubsection{Gender Perceptions and Aspirations}
Few studies showed that changes in body perception can influence gender identity and related behaviors to varying degrees (\cite{tacikowski_fluidity_2020}, but see also~\cite{Bolt2021,Provenzano2023}). In the context of the footsteps illusion, the different sound conditions were found to shift whether participants felt more feminine/masculine, as well as closer/farther to a group of women~\cite{tajadura-jimenez_as_2019, clausen_action_2021}. We here aim to replicate the \marte{transient changes in gender identity induced by acoustic corporeal cues}. \citet{tajadura-jimenez_as_2019} also found that the illusion's impact was influenced by participants' masculinity/femininity aspirations (i.e., their wish to be more masculine or more femininity), but not the participants' reported gender. A study by ~\citet{clausen_action_2021} hypothesized that the footsteps illusion would cause participants' gait to resemble masculine walking patterns more closely in the lower frequency spectra sound condition and feminine ones more closely in the high frequency spectra sound condition. Indeed, stereotypical masculine and feminine gait are known to differ in lateral hip and chest sway~\cite{Mather1994}. However, due to insufficient data, the hypothesis was not tested and is hereby explored. 
\subsubsection{Imagery Vividness and Sensory Sensitivity}
\marte{Sensory imagery has been linked to the ongoing perception of real stimuli~\cite{Imagination2020,Johansson1973,MILLER2013140}. Imaginary percepts integrate with real stimuli to conform robust multisensory experiences~\cite{berger2013}, and motor imagery involves the same neural pathways as those involved during action execution~\cite{Moran2011} and observation~\cite{Calvo-Merino2005}. Motor imagery involves similar electrophysiological patterns as illusory body ownership~\cite{EVANS2013} and is sufficient to generate a degree of illusory ownership~\cite{Berger2023}. Here, we set to investigate whether sensory imagery across modalities might modulate illusory weight changes across the different acoustic manipulations, for indeed mental imagery is also a varying individual trait potentially influencing body perception.}

\begin{table*}[t]
\begin{tabular}{lllll}
\hline
\scriptsize

       & LOW IPAQ-LOW EDEQ & LOW IPAQ-HIGH EDEQ & HIGH IPAQ-LOW EDEQ& HIGH IPAQ-HIGH EDEQ
       \\
   
  \hline
Age         & 24.05(6.26)   & 23.81(4.05)    & 25.85(8.38)   & 26.57(9.28)    \\
\hline
Weight (kg) & 68.16(12.79)  & 63.62(10.71)   & 66.05(8.21)   & 65.19(11.01)       \\ \hline
Hips (cm)   & 85.50(18.63)  & 89.86(16.36)   & 98.32(23.71)  & 90.57(12.02)    \\ \hline
Waist (cm)  & 73.18(18.98)  & 73.48(11.67)   & 75.64(9.12)   & 75.93(8.78)     \\ \hline
Height (cm) & 170.9(8.29)   & 167.71(8.86)   & 170.57(8.38)  & 170.19(8.24)    \\ \hline
EDEQ        & 0.52(0.32)    & 2.09(0.74)     & 0.56(0.35)    & 1.96(0.69)     \\
\hline
IPAQ        & \amar{1885(759)}    & \amar{2251(673)}     & \amar{6579(2400)}    & \amar{6358(2315)}   \\ \hline
\end{tabular}

\caption{Mean(SD) of weight, Hips, Waist, Height, EDEQ score and IPAQ score for the participants in four groups of 21 each. \ana{See Supplementary Material for group assignment criteria.}}
\label{fig:table_participants}
\end{table*}

\section{Methods}

\subsection{Approach and Hypotheses}
In this study, \marte{participants} wore a shoe-based prototype that modifies the frequency spectra of their walking sounds. The study was set to test the following three main hypotheses, as well as two additional secondary hypotheses:

\textit{Hypothesis 1 (Main): } \ana{the effects of} sound on body perception and feelings, emotional \ana{responses}, gait and physiology were expected to vary with subjects’ level of SED. \marte{Readers should note that we hereof refer to LOW and HIGH levels of SED according to our literature-based grouping criteria for the general population (see Supplementary Materials) and not according to clinical standards.} Participants with high levels of SED were expected to perceive their bodies heavier after hearing both the low and high frequency modified versions of their footstep sounds compared to the control condition, suggesting an impairment in the integration of external bodily signals in this group.

\textit{Hypothesis 2 (Main): } \ana{the effects of} sound on body perception and feelings, emotional \ana{responses}, gait and physiology were hypothesized to change with subjects’  PA levels. Those with LOW PA levels were expected to show larger differences between the Low Frequency (LF) and High Frequency (HF) versions of their footstep sounds, suggesting a larger malleability of body perception.

\textit{Hypothesis 3 (Main): } \ana{the effects of} sound were hypothesized to be modulated by individual body concerns. We hypothesized that people with strong negative beliefs about their body would rely more on prior expectations and therefore be more sensitive to what is generally considered positive feedback, therefore more changes for the LF condition than the HF one were expected. 

\textit{Hypothesis 4 (Secondary):} participants' reported gender were expected to be more feminine in the HF and more masculine in the LF condition~\cite{tajadura-jimenez_as_2019,clausen_action_2021}. Accordingly, we hypothesized that participants' gait patterns would be in line with such changes. 

\textit{Hypothesis 5 (Secondary):} \ana{the effects of} sound were expected to vary with subjects’ capacities for sensory imagery (i.e., imagery vividness). We hypothesized that participants with higher degrees of sensory imagery would react more strongly to the illusion.
\subsection{Participants}
84 healthy participants (mean age $\pm$ SD: 25.08 $\pm$ 7.38 years, range: 18-53; see Table ~\ref{fig:table_participants} and Supplementary Material), naïve to the study aim were recruited through the Universidad Carlos III de Madrid's participant pool, public advertisements, social networking sites, and flyers. They were pre-screened for SED by means of the Spanish version of the Eating Disorder Examination Questionnaire (EDEQ)~\cite{edeq-esp} and for PA levels using the Spanish version of the International Physical Activity Questionnaire (IPAQ)~\cite{ipaq}, \amar{using a group classification according to~\cite{IPAQSjostrom2005}}. Our objective was to create four equally-sized groups (LOW IPAQ-LOW EDEQ; LOW IPAQ-HIGH EDEQ; HIGH IPAQ-LOW EDEQ; HIGH IPAQ-HIGH EDEQ), based on the predetermined study criteria (see \ana{Table \ref{fig:table_participants} and} Supplementary Material for inclusion/exclusion, pre-screening and group assignment criteria). Participants meeting the criteria for a particular group were invited to participate, provided that the group had not reached its maximum capacity. \ana{Note that LOW and HIGH IPAQ groups differed significantly in their IPAQ scores (t=9.85, p<0.001) and that LOW and HIGH EDEQ groups differed also significantly in their EDEQ scores (t=-9.61, p<0.001)}. 

In addition to EDEQ and IPAQ scores, \ana{which directly served to test Hypotheses 1 and 2}, participants provided the following information during recruitment: demographics, including age, sex (man, woman, non-binary, decline to state), gender (masculine, feminine, non-binary person, decline to state); \amar{gender wish, ideal weight, clothes size, foot size)}; education and professional experience; and anthropometric measurements (height, weight, waist and hip circumferences). The latter was used to select the appropriate size for them for the motion suit used in the study. They also filled in the Spanish versions of the following questionnaires: the Multidimensional Body Self Relations Questionnaire (MBSRQ)~\cite{LizanaCaldern2022}, which assesses individuals’ self-attitudinal aspects towards their physical body, relates to negative/positive body image or body concerns, \ana{and which served to test Hypothesis 3}; the Betts' Questionnaire upon Mental Imagery~\cite{Campos}, which assesses the vividness with which the subject can imagine a particular experience across different sensory modalities (visual, auditory, cutaneous, kinaesthetic, gustatory, olfactory and organic), \ana{and which served to test Hypothesis 5}; and the Questionnaire on Social Support Networks (\cite{socialsupport2019}; not analyzed in this paper) (see section~\ref{sec:measures}).  
The study was conducted in accordance with the ethical standards laid down in the 1964 Declaration of Helsinki and its later amendments and was approved by the Committee of Ethics in Research of Universidad Carlos III de Madrid. All participants provided informed consent before participating and received a €20 bank transfer as compensation for their time.

\begin{figure*}[ht]
\includegraphics[width=15.5cm]{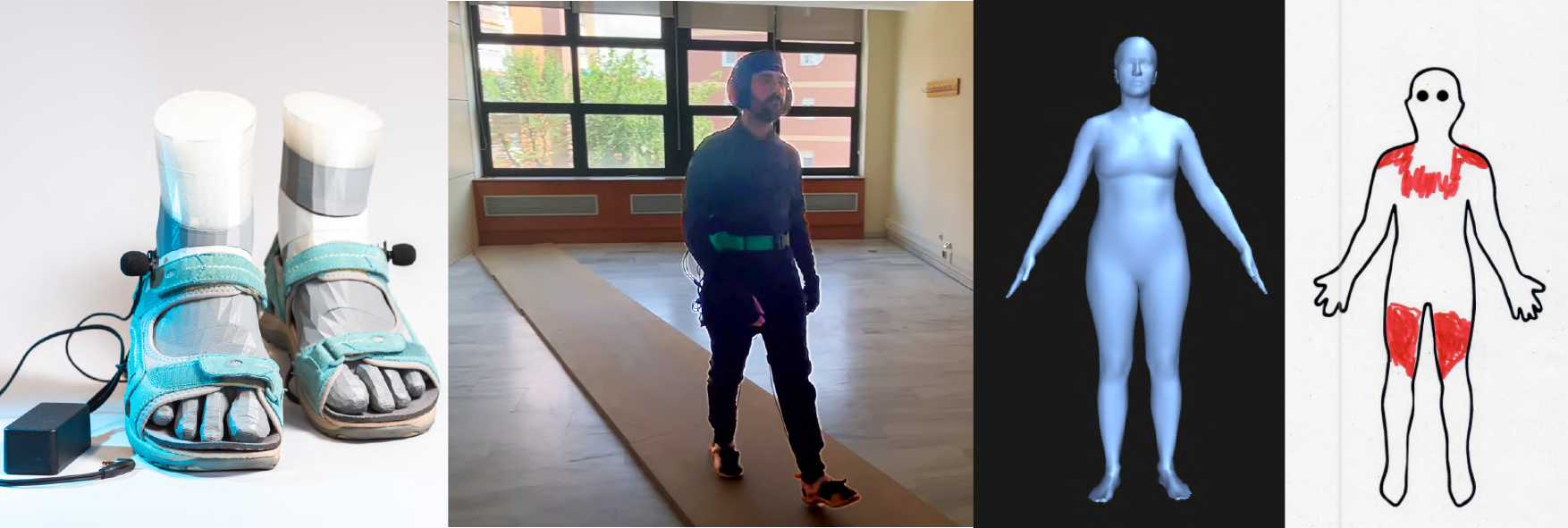}

\caption{Detail of the microphones in the prototype; laboratory setup and participant wearing prototype and mocap suit; Body Visualizer tool employed to measure visualized body size/weight; an example of a body map filled by participants.}
\label{fig:methods_figure}
\Description{Four adjacent pictures showing: The microphone placing in SoniWieght Shoes; the laboratory setup with a participant wearing prototype and mocap suit; a screenshot from the body visualizer tool employed to measure visualized body size/weight; an example of a body map filled by a participant}
\end{figure*}

\subsection{Trial Design}
\subsubsection{Apparatus and Materials}
 We used \amar{SoniWeight Shoes}, a real-time footstep sound modification system adapted from related studies~\cite{tajadura-jimenez_as_2015, tajadura-jimenez_as_2019, tajadura-jimenez_bodily_2017, brianza_as_2019, clausen_action_2021}. This version prioritizes portability and ergonomics for enhanced comfort, reduced data loss and extended walking distances. Described in detail in~\cite{tecniacustica_delaprida2022} \ana{and in Supplementary Material}, the system comprises strap sandals with attached microphones (Core Sound) for capturing footstep sounds. Unlike earlier versions~\cite{tajadura-jimenez_as_2015, tajadura-jimenez_as_2019, tajadura-jimenez_bodily_2017, brianza_as_2019, clausen_action_2021}, our system incorporates a DSP electronic board (Bela.io) to amplify and modify the sound spectrum. Three sound feedback conditions were created, namely “High Frequency” (HF), amplifying 1–4 kHz components and attenuating 83–250 Hz by 12 dB; “Low Frequency” (LF), amplifying 83–250 Hz and attenuating above 1 kHz by 12 dB; and “Control” (C), maintaining natural footsteps equally amplified across frequency bands, i.e., with flat frequency response. These conditions mirror those in prior studies~\cite{tajadura-jimenez_as_2015,brianza_as_2019} (see also the study by~\cite{li1991}, in which the spectral components of waking sounds were manipulated similarly to influence the perception of another person's body characteristics). Modified sounds were presented through headphones (Sennheiser HDA300) with high passive ambient noise attenuation (>30 dBA). Since the ground material and type of footwear are relevant for the footstep sounds, wooden pavement and two pairs of strap sandals (EU size 37 and 42) were chosen. The latter are easy to wear and have a hard rubber sole that elicits clear and distinctive footstep sounds.

A Rokoko smart suit, equipped with 19 inertial measurement units (IMU) featuring 9 degrees of freedom (DoF) sensors, and a 100 Hz resolution, measured positional triplets for 17 anatomical points in 3D Cartesian space. Four suit sizes (S/M/L/XL) accommodated all participants. They were asked to wear a cotton T-shirt and trousers for comfort. 
A Bitalino PLUX acquisition system (8-Channel biosignalsplux hub) with 1000 Hz sampling resolution captured physiological signals including electromyogram (EMG), electrocardiogram (ECG), respiration (RESP), electrodermal activity (EDA) and acceleration (ACC). Notably, respiration data were excluded from this study due to noise. \amar{A custom GUI was developed to connect and calibrate the two acquisition devices, define filenames and start recordings, synchronously managing requests, awaiting responses and compiling logs and timestamps in text files.}
\subsection{Experimental Design}
A mixed experimental design involved participants from four experimental groups undergoing two repetitions of three sound conditions (HF, LF, C) in a randomized order. Two between-subjects independent variables were the EDEQ and IPAQ levels, determined by prior demographic assessments. Dependent outcome measures, recorded either during (physiological and motor) or after (questionnaires) the experimental manipulations, are presented below.  
\amar{Sensor locations on participants' body are shown in Figure \ref{fig:teaser}b).}

\subsubsection{Outcome Measures}
\label{sec:measures}
To evaluate the effects of sound condition on participants' \ana{bodily and emotional} experience, we combined self-reports, behavioral and physiological measures, as described below. \ana{Collectively, these measures were employed to examine the five hypotheses presented earlier. We refer to emotional responses as short-time processes in response to \ana{stimuli \cite{damasio2006descartes}}, in this case the experimental condition, and which can be measured by looking at subjective reports, physiological reactivity and overt behavioral acts~\cite{BRADLEY}. Here we look at \ana{the subjective experience of emotions \cite{damasio2006descartes}}, which we characterize in terms of valence, arousal, and dominance in line with dimensional theories of emotion (e.g.~\cite{Russell1980,Lang1995TheEP}), and physiological responses induced by sound}.

\textbf{Questionnaire on Body \ana{feelings} and Emotional \ana{experience}}: We quantified body feelings using a Likert-type questionnaire (see Supplementary Material), adapted from~\cite{tajadura-jimenez_as_2015,tajadura-jimenez_as_2019}. Comprising 9 statements, the first 5 items assessed the felt walking speed (ranging from slow to quick), body weight (ranging from light to heavy), body strength (ranging from weak to strong), body straightness (ranging from stooped/crouched to elongated/extended) and the sensation of being more feminine/masculine. The next 4 items assessed the agreement levels (from strongly disagree to strongly agree) with statements, including one checking whether participants felt agency over the sounds, as many studies have shown that large discrepancies between modalities and delays between actions and sensory feedback disrupt agency and diminish the sensory-induced bodily illusions~\cite{menzer_feeling_2010, tajadura-jimenez_action_2012}. Additional statements assessed whether participants had vivid (vividness) of unexpected feelings about their body (surprise)~\cite{tajadura-jimenez_action_2012}, and feelings on feet localization, as sound may interact with proprioception~\cite{tajadura-jimenez_action_2012}. Emotional \ana{experience} was assessed using the self-assessment manikin questionnaire~\cite{BRADLEY}, consisting of three 9-item graphic scales of valence, arousal and dominance. \ana{This questionnaire, widely used in assessing emotional responses to acoustic stimuli~\cite{BRADLEY2000}, has been applied in prior studies on the footsteps illusion~\cite{tajadura-jimenez_as_2015, tajadura-jimenez_as_2019, tajadura-jimenez_bodily_2017}}.

\textbf{Body Visualization}: To collect participants' visual estimates of their own body weight, we used the Body Visualizer tool~\cite{Bodyvisualizer}, employed in similar studies for the same purpose~\cite{tajadura-jimenez_as_2015,tajadura-jimenez_as_2019,tajadura-jimenez_bodily_2017,tajadura-jimenez_body_2022,Piryankova_weight}. Following each experimental trial, participants performed the task twice. The experimenter set the height of the avatar to match the participants' and set the initial weight to 75\% or the 125\% of the participants' actual weight. Using a trackpad, participants then adjusted the 'weight' dimension of the avatar to reflect their own perceived body size (see similar procedures~\cite{tajadura-jimenez_as_2015,tajadura-jimenez_as_2019,tajadura-jimenez_body_2022}). The order of the initial weight (75\% or 125\%) was counterbalanced across two subsequent repetitions. The analysis considered the average from the two trials for each condition.

\label{sec:biomecanic_measures}
\textbf{Gait Biomechanics}: These were used as an implicit measure of changes in body perception, following previous studies~\cite{tajadura-jimenez_as_2015,tajadura-jimenez_as_2019} suggesting that the sound-driven illusion of altered body weight results in people adapting their gait to the prototypical motor pattern of lighter/heavier bodies. As in those studies, we quantified leg lifting acceleration and time parameters for each gait cycle (i.e., the time between two successive steps made by one foot)~\cite{cunado2003automatic}, including stance and swing phase times. Gait patterns are characterized by larger leg accelerations for people with lighter bodies and slower gait speed and a longer duration of the stance and the heel strike for people with heavier bodies~\cite{ko2010characteristic,troje2008retrieving}. 
Knee flexion data from motion suit recordings were used. Angles were plotted in MATLAB, and peaks and valleys were extracted to determine step count and duration over the various task phases (e.g., walk-in-place, first round along the corridor, and return phase). This data extraction from each sound condition was used for statistical analyses (further details in Supplementary Material). In addition, IMUs captured lateral hip sway (left hip's external rotation and flexion) and chest sway (shoulders' and thorax movements)~\cite{clausen_action_2021}, to investigate Hypothesis 4 related to how movements correlate with changes in gender identity.
Leg and shoulder muscular activity was recorded using EMG sensors on the  vastus medialis oblique and cervical trapezius sites, as shown in fig \ref{fig:teaser}b). 

\textbf{Physiological responses}: We measured EDA, linked to emotional arousal and stressful states (emotional sweating)~\cite{boucsein2012electrodermal}, \ana{and its association with the footsteps illusion has been established~\cite{tajadura-jimenez_as_2015}}. ECG was also measured, serving as a real-time measure of attention and arousal~\cite{Andreassi,Dillon2000}. Specifically, increased attention results in short-term heart rate deceleration, while arousal induces long-term acceleration of the heartbeat~\cite{Lang1990}. Heart rate variability (HRV) has also been connected to mental social cognition processes: increased HRV may serve as a marker for one's ability to recognize emotions in humans~\cite{Quintana2012}. For the HRV analyses performed in this study, refer to~\cite{Shaffer2017AnOO}.
Heart rate can also indicate the valence of an emotional response, discriminating between positive and negative emotions~\cite{Bolls2001,Cuthbert1996,Greenwald1989,Lang1990}. 

\label{sec:procedure}
\begin{figure*}[t] 
\centering
\includegraphics[width=\linewidth]{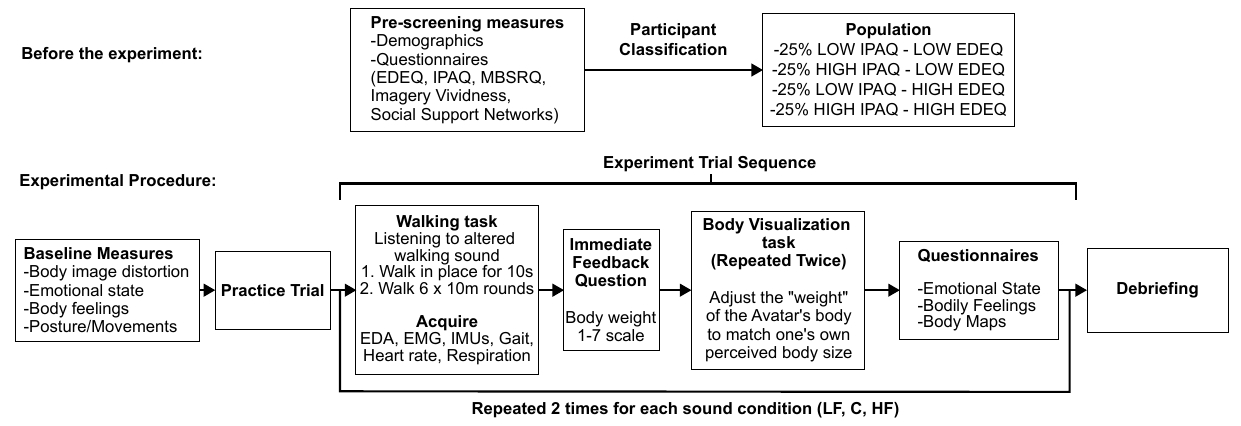}
\caption{Experiment procedure}
\label{fig:procedure}
\Description{Diagram composed of 2 main sections: the first composed of 2 blocks: the first is Pre-screening measures and the second describes Population composition. The second section is composed of 7 blocks starting from Baseline Measures and ending with Debriefing, details provided in Procedure section}
\end{figure*}

\textbf{Body Maps}:
Body maps, a prevalent tool in psychology and HCI research for capturing subjective accounts of body experiences~\cite{turmo_vidal_towards_2023}, traditionally depict a human body silhouette in its frontal plane where individuals express experienced sensations and emotions through drawings, scribbles, and/or symbols. Within HCI, body maps are often used for qualitative data gathering (see~\cite{anne_cochrane_body_2022, turmo_vidal_towards_2023} for overviews). Here, drawing inspiration from psychology research~\cite{Nummenmaa,Suvilehto}, we innovatively applied body maps as a quantitative data gathering tool. We employed them to capture changes felt by participants in their body sensations or perceptions - with the aim of quantifying them and exploring any  differences according to individual factors and sound conditions. Participants marked changes in body sensations or perceptions on paper body maps after each sound condition and repetition (i.e., two body maps for each sound condition). Based on an initial open-coding of participants' responses to the body maps, we divided it into 136 individual regions across the front and back maps (e.g., right thumb finger in the front body map, left index finger in the back body map), which were clustered into thematic groups (e.g., hand) all the way up to broad body regions encompassing limbs, head and trunk (e.g., left arm), see Supplementary Material. We classified each participant's body map according to these areas, and summed the regions according to sound conditions, facilitating quantitative analysis of perceived changes.

\subsection{Procedure}

Figure~\ref{fig:procedure} shows a graphical overview of the experimental procedure. Participants were categorized based on their SED and PA levels during pre-screening. Baseline measures and the experimental manipulation were performed in a single session. The procedure involved 2 repetitions of each of the 3 acoustic manipulations (LF, HF and C). In each trial, participants were asked to walk in a fixed position for 10 seconds and then back and forth in a straight line for 6 rounds at a comfortable speed; after each round they were asked to report their felt body weight, using a 7-point Likert scale ranging from light to heavy (immediate feedback question). Physiological and behavioral measures were taken during walking. Upon completion of each walk, participants completed the body visualization task and questionnaires on a computer. The session concluded with a short debriefing with the experimenter.

\subsection{Statistical Analyses and Data Treatment}
Physiological data were processed in Matlab (version R2022a) using band-pass and low-pass Butterworth filters for ECG, EDA, and EMG signals according to Bitalino's toolkit. The resulting data were exported for statistical analyses. Rokoko's mocap suit extraction details are provided in~\ref{sec:biomecanic_measures} (\textit{Gait Biomechanics}). Statistical analyses were computed in R (version 4.3.1, R Core Team, 2018). Alpha levels were set to 0.05, and p-values were adjusted for multiple comparisons. Data normality was assessed through visual inspection. For non-parametric comparisons (i.e., questionnaires), aligned rank transform ANOVAs (ART ANOVA~\cite{Wobbrock}) were computed because of their robustness ~\cite{Wobbrock2011}. Significant findings for these comparisons were followed up by Wilcoxon signed-rank tests, \amar{adjusted for multiple comparisons using Bonferroni correction}. Comparisons for parametric data involved mixed model ANOVAs with follow-up T-tests for the comparisons of interest, \amar{also applying a Bonferroni correction for multiple comparisons}.  Further inspection of covariates was computed using ANCOVA, which has been reported to be robust to deviations from normality~\cite{McDonald2009}.

\subsection{Dataset}
We provide a dataset including participants' demographic data (age, sex, gender, gender wish, height, weight, ideal weight, waist, hips, clothes size and foot size); survey data (EDEQ, IPAQ, MBSRQ, sensory imagery, social support networks, body \ana{feelings and emotional experience}), behavioral data (avatar task values, average values from the body movement and data from sensors), and physiological data (average values from EMG, ECG, RESP, EDA and IMU sensor data). This ensures transparency and encourages researchers to use it for further analysis, study comparisons or reviews. Data is integrated into a CSV file with measures for each sound condition and repetition, anonymized with numerical IDs. Metadata is used to make the dataset findable and provide comprehensive insights such as the description of the project, the equipment used, details about the variables, and other critical considerations for users.

\begin{table*}[ht!]
\small
\caption{Effects of Sound Condition for questionnaire data (7-level Likert items except for 9-level valence, arousal and dominance scales). Median(Range) values \marte{for the average of both repetitions} and results of ANOVAs on aligned-ranked data and \marte{Bonferroni-corrected} Wilcoxon Pairwise Comparisons are shown.
}
\begin{tabular}{llllll}
\hline
     & & \begin{tabular}[c]{@{}l@{}}ANOVA on aligned-rank\\ transformed data\end{tabular} & \multicolumn{3}{l}{Wilcoxon Pairwise Comparisons} \\ \hline

\textbf{Measure}         & \textbf{C   HF   LF} & \textbf{Effect of Sound Condition} & \textbf{High vs Low} & \textbf{High vs Control} & \textbf{Low vs Control} \\ \hline
\textbf{Quickness}      & 4(6) 5(6) 4(6)  & \textbf{(F=11.22, p<0.001, $\eta_{p}^{2}=0.12)$} & \textbf{z=-4.75 p<0.001} & \textbf{z=-2.72 p=0.019} & \textbf{z=-2.40 p=0.048}\\ \hline
\textbf{Weight}         & 3(6) 3(6)  4(6)  & \textbf{(F=20.10, p<0.001, $\eta_{p}^{2}=0.20)$} & \textbf{z=-5.83 p<0.001} & \textbf{z=-2.48 p=0.039} & \textbf{z=-4.72 p<0.001}\\ \hline
\textbf{Strength}       & 4(6) 4(6) 4(6) & \textbf{(F=3.65, p=0.028, $\eta_{p}^{2}=0.04)$}  & z=-1.29 p=0.58 & \textbf{z=-3.07 p=0.006} & z=-1.13 p=0.77\\ \hline
\textbf{Straightness}     & 5(6) 5(6) 5(6) & \textbf{(F=3.66, p=0.027, $\eta_{p}^{2}=0.04)$} & \textbf{z=-2.89 p=0.011} & z=-1.22 p=0.66 & z=-2.15 p=0.093 \\ \hline
\textbf{Masculinity}      & 4(6) 4(6) 4(6)  & \textbf{(F=10.02,p<0.001, $\eta_{p}^{2}=0.11)$} & \textbf{z=-3.99 p<0.001} & z=-1.11 p=0.79 & \textbf{z=-3.48 p=0.001}\\ \hline
\textbf{Proprioception} & 6(6) 5(6) 5(6) & \textbf{(F=15.31, p<0.001, $\eta_{p}^{2}=0.15)$} & z=-0.26 p=1 & \textbf{z=-5.30 p<0.001} & \textbf{z=-4.84 p<0.001} \\ \hline
Vividness      & 3(6) 3(6) 3(6) & (F=1.86, p=0.16, $\eta_{p}^{2}=0.02)$ & z=-1.22 p=0.66 & z=-0.92 p=1 & \textbf{z=-2.46 p=0.040}\\ \hline
\textbf{Surprise}     & 3(6) 4(6) 4(6) & \textbf{(F=8.77, p<0.001, $\eta_{p}^{2}=0.10)$}  & z=-1.63 p=0.30 & \textbf{z=-4.13 p<0.001} & \textbf{z=-3.27 p=0.003}\\ \hline
\textbf{Agency}  & 6(6) 6(6) 6(6) & \textbf{(F=4.09, p=0.018, $\eta_{p}^{2}=0.05)$} & z=-0.29 p=1 & z=-1.64 p=0.29 & z=-1.89 p=0.17\\ \hline
\textbf{Valence}      & 7(7) 7(8) 6(8)  & \textbf{(F=4.49, p=0.013, $\eta_{p}^{2}=0.05)$}  & \textbf{z=-2.98 p=0.008} & z=-1.56 p=0.35 & z=-1.86 p=0.18\\ \hline
Arousal        & 3(8) 3(7) 3(7) & (F=1.00, p= 0.37, $\eta_{p}^{2}=0.01)$ & z=-1.22 p=0.66 & z=-1.65 p=0.29 & z=-0.37 p=1 \\ \hline
Dominance      & 5(8) 5(8) 5(8) & (F=0.96, p= 0.38, $\eta_{p}^{2}=0.01)$ & z=-0.29 p=1 & z=-0.48 p=1 & z=-1.20 p=0.68 \\ \hline
\end{tabular}
\label{fig:quest_table}
\end{table*}

\section{Results}

We report the statistical results for the overall population, followed by how each of the targeted individual factors interacts with the effects (according to SED, PA level, body concerns and sensory imagery). For all analyses, we report the effect of repetition only when significant effects were found for it or its interaction with the different sound conditions. \ana{If no effect of repetition or interaction was found we report the average of both trials for each sound condition}.

\subsection{Effects on the Overall Population}

\subsubsection{Questionnaire on Body \ana{Feelings and Emotional Experience}}
\begin{figure*}[ht]
\includegraphics[width=13.2cm]{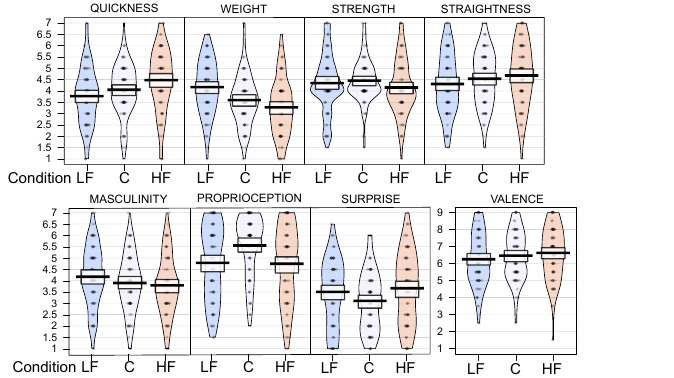}
\caption{Questionnaire results in overall population}
\label{fig:quest_results}
\Description{Boxplots of the questionnaire items: for details of the values that are shown, refer to Table 2.}
\end{figure*}

ART ANOVAs on the questionnaire data revealed a main effect of sound condition for several questionnaire items. As shown in Table~\ref{fig:quest_table} and Figure~\ref{fig:quest_results}, there were significant main effects of sound condition on quickness, weight, strength, straightness, masculinity, proprioception, surprise, agency and emotional valence. Participants felt quicker in the HF than both in LF and C, as well as in C vs LF; lighter in the HF than in both LF and C conditions as well as in C vs LF. They reported feeling heavier in LF than in both HF and C conditions, as well as in LF vs C. Stronger in C than in H. Higher masculinity was found for the LF than the HF condition, and in LF than in C. In HF participants felt they were standing more straight up and felt happier than in LF. Participants stated that in both the frequency manipulated conditions they were more surprised with the sounds they heard, while at the same time found more difficult to localize their feet, than in C. Importantly, while for agency (i.e., the feeling of one producing the sounds) there was a main effect, follow-up analyses did not show significant differences between conditions, with participants reporting high agency feelings for all sounds. It was important to confirm that agency was preserved across conditions, as previous studies have shown that disrupting agency can diminish the sensory-induced body illusions~\cite{menzer_feeling_2010, tajadura-jimenez_action_2012}. 
\amar{Note that, as per~\cite{Cohen1988}, the effect size of sound condition is large for weight and proprioception, and medium for quickness, masculinity and surprise. However, the effect size for strength, straightness, agency and valence is small; hence, these findings should be interpreted cautiously (though see~\cite{Morris2020MisunderstandingsAO}, recognizing the potential significance of small effect sizes). }

\subsubsection{Body Visualization}
ANOVA results showed a significant effect of sound condition on the weight of the visualized body (F=9.52, p<0.001, $\eta_{p}^{2}=0.10$), \ana{being this effect size medium~\cite{Cohen1988},} pairwise comparisons between sound conditions revealed that participants represented their body as slimmer in HF compared to LF (z=-5.18, p<0.001) (Fig.~\ref{fig:overall_results}).

\subsubsection{Gait Biomechanics}

\begin{figure*}[ht]
\includegraphics[width=12cm]{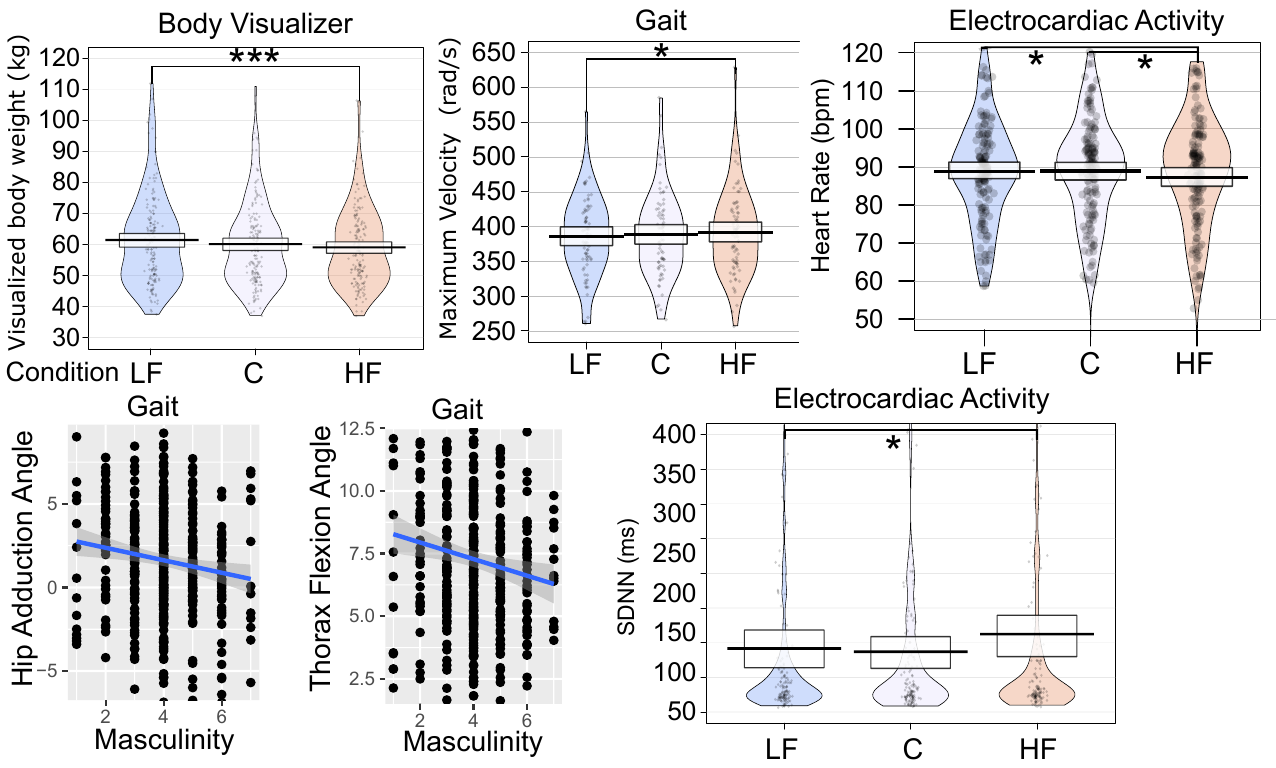}
\caption{Main results showing effects of sound condition on body visualization, gait biomechanics and physiology in the overall population. Correlations between hip and thorax sway and participants' reported feelings of masculinity induced by sound are also displayed.}
\Description{Picture composed of a series of 7 body maps (front and rear body each) showing the significant differences that were found in the ANOVAs respectively for: Sound condition, EDEQ groups, IPAQ groups, and the combination of sound and EDEQ and IPAQ groups. The body maps areas where significant differences were found are painted.}
\label{fig:overall_results}
\end{figure*}

\textbf{Leg Movement} Gait IMU data was lost for 8 participants.  An analysis of the remaining data showed a significant effect of sound condition on maximum step velocity (F=3.24, p=0.042, $\eta_{p}^{2}=0.05$); \amar{however, the size of this effect is quite small~\cite{Cohen1988}}. Participants moved faster in HF compared to LF (p=0.032), as shown in Figure~\ref{fig:overall_results}. Means~($\pm$~SD) were for LF: 369~($\pm$~76), for HF: 376~($\pm$~74) and for C: 378~($\pm$~73). A double interaction between sound condition and repetition was found for maximum acceleration (F=4.26, p=0.016, $\eta_{p}^{2}=0.05$); separate analyses for each repetition yielded only for repetition 2 a significant effect of sound condition (F=3.74, p=0.026, $\eta_{p}^{2}=0.06$), \ana{with a medium effect size~\cite{Cohen1988}}. Participants accelerated more their legs in HF contrasted with LF (p=0.049). Means~($\pm$~SD) were for LF: 3988~($\pm$~937), for HF: 4067~($\pm$~861) and for Control: 4078~($\pm$~873).

\textbf{Leg Muscular Activations} 
A significant effect of sound condition was found on EMG muscle activation  (F=3.23, p=0.042, $\eta_{p}^{2}=0.04$); \ana{however, the effect size was small~\cite{Cohen1988}}. As Figure~\ref{fig:overall_results} shows, participants showed higher values of EMG muscle activation in HF compared to LF (p=0.028). Means~($\pm$~SD) were for LF: 57.7~($\pm$~26.9), for HF: 60.4~($\pm$~31.0) and for C: 62.1~($\pm$~32.2). 

\textbf{Lateral Hip and Shoulder Sway}
ANOVAs performed on the hip flexion angle mean provided significant differences across sound conditions (F=3.61, p=0.029, $\eta_{p}^{2}=0.05$); \ana{however, the effect size was small~\cite{Cohen1988}}. Participants moved with higher values of hip flexion angle on LF in comparison with C (p=0.020). Means~($\pm$~SD) were for LF: 39.12~($\pm$~6.63), for HF: 38.79~($\pm$~6.16) and for C: 38.33~($\pm$~5.86). 
Additional analyses were computed to test the hypothesis related to participants' gender perception and aspirations, with the values provided by them prior to the experiment. A significant interaction, \ana{with a medium effect size~\cite{Cohen1988}}, was found between how feminine/masculine reported they perceived themselves to be and hip flexion angle (F=6.62, p=0.012, $\eta_{p}^{2}=0.09$), as shown in Figure~\ref{fig:overall_results}.
To further explore our hypothesis 4 in the interrelation of gender perception and movement, we ran two additional Spearman's rank correlation analyses. We correlated lateral hip and shoulder sway data in relation to masculinity answers from the post-condition questionnaires. A significant correlation was found between hips adduction angle and masculinity feelings (p=0.006, \amar{r=-0.133}), as well as between thorax lateral flexion rotation angles and masculinity feelings (p=0.005, \amar{r=-0.134}). As shown in fig ~\ref{fig:overall_results}, for higher values of masculine feelings, both hips and thorax angles values reduce, according to~\cite{clausen_action_2021}.
\subsubsection{Physiological Responses}

\textbf{Electrodermal Activity} No significant overall effects were found for this measure.

\textbf{Electrocardiac Activity}
ANOVAs yielded a significant effect across sound conditions for heart rate (F=9.45, p<0.001, $\eta_{p}^{2}=0.11$) during the walk in place phase of the trial. For the whole experimental trial, participants have lower values of heart rate in HF compared to LF (p=0.043) and to C (p=0.025). Means~($\pm$~SD) were for LF: 89.22~($\pm$~13.40), for HF: 87.38~($\pm$~15.67) and for C: 88.78~($\pm$~14.15).
The standard deviation of NN intervals (SDNN) showed significant differences across conditions (F=5.08, p=0.007, $\eta_{p}^{2}=0.06$). Participants had higher values of SDNN in HF compared to LF (p=0.043). Means~($\pm$~SD) were for LF: 91.52~($\pm$~167.87), for HF: 112.33~($\pm$~195.17) and for C: 86.76~($\pm$~149.68). Root mean square of successive RR interval difference (RMSSD) also found significant differences across conditions (F=4.12, p=0.018, $\eta_{p}^{2}=0.5$). Participants have higher values of RMSSD in HF compared to LF (p=0.022) and C (p=0.032). Means~($\pm$~SD) were for LF: 94.87~($\pm$~203.08), for HF: 158.17~($\pm$~349.17) and for C: 100.85~($\pm$~200.69). \amar{The effect size is notably large for RMSSD, while heart rate and SDNN exhibit medium effect sizes~\cite{Cohen1988}.}

\textbf{Effects on Body Maps}
When comparing sound conditions, ANOVAs showed that participants felt more changes in the lower body for LF, while HF provoked more effects in the upper body (inner arms, shoulders).
For detailed statistics of body maps areas see Supplementary Results. In Figure~\ref{fig:bmaps_results}, we report the differences we found through this analysis.

\begin{figure*}[ht]
\includegraphics[width=17cm]{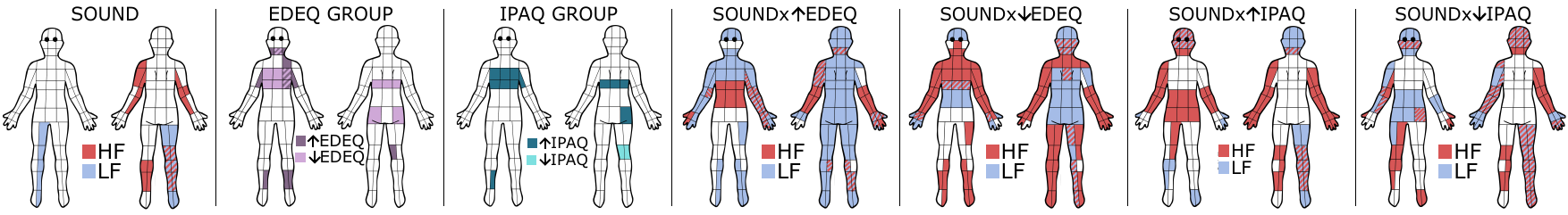}

\caption{Effects showing the main differential effects of sound, EDEQ and IPAQ groups and effects of sound combined to differences in EDEQ and IPAQ on the various areas of the Body Maps}
\label{fig:bmaps_results}
\Description{Picture composed of a series of 6 graphs: three boxplots showing means of Body Visualizer reported weight, Maximum velocity and heart rate for the three sound conditions. Two plots are then showing hip adduction and thorax flexion correlation with masculinity values reported in questionnaires. Lastly, a boxplot shows SDNN ECG feature variation with sound condition. }
\end{figure*}

\subsection{Individual Differences in the Effects}
The overall effects exposed more particular interconnections when including individual differences in levels of EDEQ and IPAQ, as well as body concerns and sensory imagery, as explained below. We report results only for those measures for which we found significant interaction effects between the investigated individual score and sound condition. 

\subsubsection{Differences According to Eating Disorder Symptomatology (EDEQ)}

\begin{figure*}[ht]
\includegraphics[width=14cm]{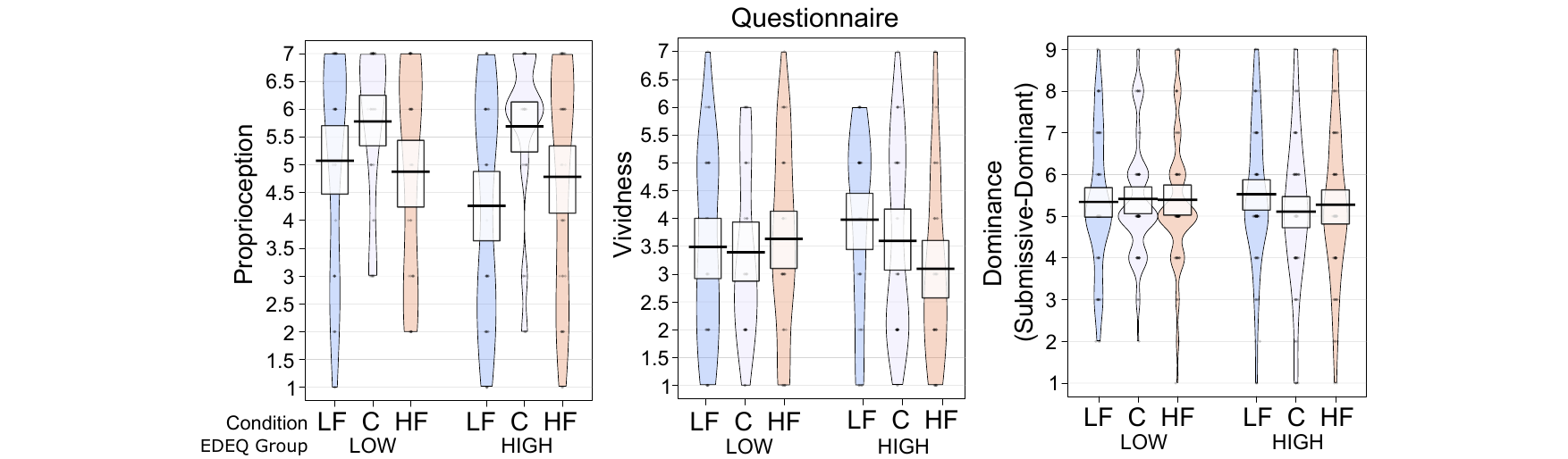}
\caption{Results showing the main differential effects of sound condition for the two different EDEQ groups on questionnaires answers for proprioception, vividness and dominance and on EDA mean value.}
\label{fig:edeq_results}
\Description{Picture composed of a series of 4 boxplots showing means of proprioception, vividness and dominance answers to the questionnaires and electrodermal activity (EDA) means across sound conditions for the two EDEQ groups}
\end{figure*}

\textbf{Questionnaire Results}
Considering the between-subjects factor EDEQ group (i.e., high vs low EDEQ score), there was a significant\amar{, yet small in effect size~\cite{Cohen1988},} interaction between sound condition and EDEQ group for dominance (F=3.17, p=0.045, $\eta_{p}^{2}=0.03$); as shown in Figure~\ref{fig:edeq_results}, differences according to sound condition were more pronounced for the HIGH EDEQ, who felt more dominant with LF than with the other two conditions. Further, there was a significant triple interaction between sound condition, repetition and EDEQ group for proprioception (F=4.26, p=0.015, $\eta_{p}^{2}=0.03$) and vividness (F=3.35, p=0.037, $\eta_{p}^{2}=0.04$). Separate analyses for each repetition showed significant effects of sound only for the first repetition for both vividness (F=3.56, p=0.031, $\eta_{p}^{2}=0.04$) and proprioception (F=2.82, p=0.049, $\eta_{p}^{2}=0.03$); \amar{yet the size of these effects is small~\cite{Cohen1988}.}  As shown in Figure~\ref{fig:edeq_results}, we observe more pronounced differences between sound conditions for the group with HIGH EDEQ, who felt larger loss in ability to localize their feet in the frequency-altered conditions vs C, as well as felt a more vivid experience of their body with LF than with the other two conditions, especially than in HF.

\textbf{Gait Biomechanics}
There was a significant interaction between sound condition and EDEQ group for mean step velocity (F=5.68, p=0.004, $\eta_{p}^{2}=0.07$) and for mean acceleration (F=7.86, p<0.001, $\eta_{p}^{2}=0.09$), \ana{both with medium effect sizes~\cite{Cohen1988}. Nevertheless, }separate analyses for each group did not show significant differences between conditions.

\textbf{Electrodermal Activity}
There was a significant interaction between sound condition and EDEQ group for mean EDA value (F=3.93, p=0.021, $\eta_{p}^{2}=0.04$), \ana{but the effect size was small~\cite{Cohen1988}}; as shown in Figure~\ref{fig:edeq_results}, differences due to sound condition were more pronounced for the LOW EDEQ group, for which EDA was lower in the frequency-altered sound conditions. The HIGH EDEQ group showed a reverse pattern, showing higher EDA values for the frequency-altered sound conditions. Separate analyses for each group did not show significant differences between conditions.

\textbf{Effects on Body Maps}
Excluding the effect of sound condition, and considering only EDEQ groups, we found some changes in the mouth area for LOW EDEQ (not affected in HIGH EDEQ), legs and inner arms, while for LOW EDEQ the bottom area and the dorsal area of the back was affected by some changes. 
Separately including the effect of sound condition, an ANOVA for the HIGH EDEQ group, showed a stronger response to LF, affecting mostly the entire rear body, upper head, mouth and chin, chest, arms and distal lower limbs. HF had a stronger effect on arms, lower chest and waist areas. Neck, hands and parts of the arms showed effects both in HF and LF. As for the LOW EDEQ group LF produced changes in the upper head, lower chest and dorsal and lumbar back, while HF affected more the upper chest, distal arms and legs. See Supplementary Results for detailed statistics and Fig.~\ref{fig:bmaps_results} for a visual depiction.

\begin{figure*}[ht]
\includegraphics[width=15cm]{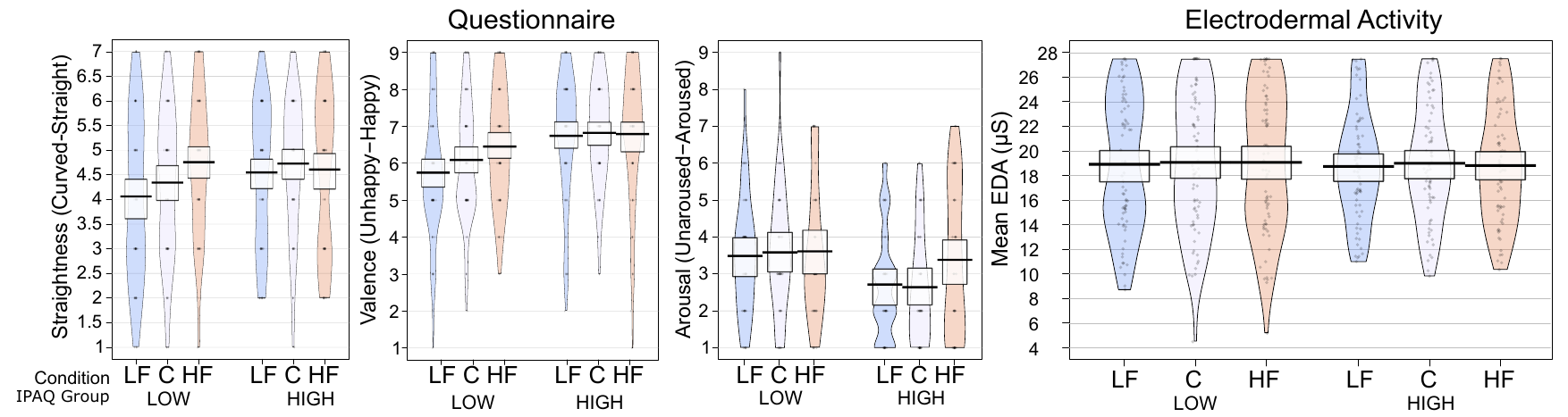}
\caption{Results showing the main differential effects of sound condition for the two groups differing in IPAQ-level.}
\Description{Picture composed of a series of 4 boxplots showing means of straightness, valence and arousal answers to the questionnaires and electrodermal activity (EDA) means across sound conditions for the two IPAQ groups}
\label{fig:ipaq_results}
\end{figure*}

\subsubsection{Differences According to Level of Physical Activity (IPAQ)} 

\textbf{Questionnaire Results}
When including the participants' IPAQ group as a factor (i.e., high vs low IPAQ score), there was a significant, \amar{yet small in effect size}, interaction between sound condition and IPAQ group for straightness (F=5.35, p=0.006, $\eta_{p}^{2}=0.01$). As shown in Figure~\ref{fig:ipaq_results}, differences in straightness due to sound condition were more pronounced for the LOW IPAQ level group, who felt straighter with HF than with the other conditions. There was also a triple interaction between sound condition, repetition and PA group for surprise (F=3.16, p=0.045, $\eta_{p}^{2}=0.01)$, but separate analyses for each repetition did not yield significant effects. There were significant interactions between sound condition and IPAQ group for valence (F=3.24, p=0.042, $\eta_{p}^{2}=0.04)$ and arousal (F=4.06, p=0.019, $\eta_{p}^{2}=0.02)$,  \ana{but the effect sizes were small}. Differences in valence due to sound condition were more pronounced for LOW IPAQ group (Figure~\ref{fig:ipaq_results}), who felt happier with HF than with the other two conditions. In contrast, while arousal was overall higher for the inactive group, differences in arousal were higher for the active group, who felt more aroused with HF than with the other conditions. There was also a triple interaction between sound condition, repetition and IPAQ group for arousal (F=3.44, p=0.034, $\eta_{p}^{2}=0.04$); separate analyses for each repetition showed a significant, \ana{yet with small effect size}, interaction between sound condition and IPAQ group only for the second repetition (F=3.79, p=0.025, $\eta_{p}^{2}=0.04$),  where more pronounced differences due to sound condition were observed.

\textbf{Electrodermal Activity}
There was a significant interaction between sound condition and IPAQ group for Mean EDA value (F=3.46, p=0.034, $\eta_{p}^{2}=0.04$), \ana{but the effect size was small~\cite{Cohen1988}}. As shown in Figure~\ref{fig:ipaq_results}, for physically inactive participants, EDA mean was \ana{lower for LF}, while for active participants, EDA reduced for both HF and LF. Separate analyses for each group did not show significant differences between conditions.

\textbf{Effects on Body Maps} Excluding the effect of sound condition, most of the changes involved in the HIGH IPAQ group, involve the whole chest, dorsal back and partially the bottom area. When including sound condition as a factor, participants with high levels of IPAQ felt more changes in HF, specifically in  the lower chest and waist, arms and hand dorsums, face, mouth and neck area. LF showed effects on the head, feet and the rear part of the right leg. For the LOW IPAQ group, results showed changes in limbs for both HF and LF, while head, lower chest and waist are more affected by LF. See Supplementary Results for statistic values and Fig.~\ref{fig:bmaps_results} for a visual depiction.

\subsubsection{Differences According to Body Concerns (MBSRQ)}

\begin{figure*}[ht]
\Description{Picture composed of a series of 5 plots showing variation of hip flexion, mean velocity, strength (questionnaire item), electrocardiac activity (ECG) and surprise (questionnaire item) for the three sound conditions varying values of MBSRQ}
\includegraphics[width=16.5cm]{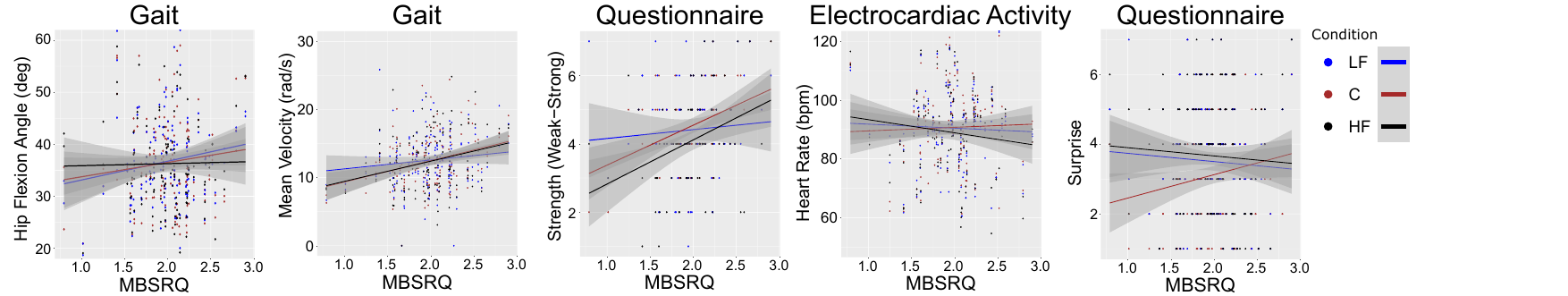}
\caption{Results showing the main differential effects of sound due to differences in body concerns (MBSRQ score used as a covariate).}
\label{fig:mbsrq_results}
\end{figure*}

\textbf{Questionnaire Results}
The interaction with different degrees of body concerns were evaluated by adding the participants' MBSRQ score as a covariate. ANCOVA results showed an interaction between sound and the MBSRQ score for surprise (F=4.05, p=0.019, $\eta_{p}^{2}=0.05$), \ana{with a small effect size}, and a triple interaction between sound, repetition and MBSRQ score for strength (F=5.12, p=0.007, $\eta_{p}^{2}=0.06$), \ana{with a medium effect size}. Separate analyses for each repetition for the variable strength showed a significant interaction between MBSRQ score and sound condition only for the first repetition (F=2.23, p=0.049, $\eta_{p}^{2}=0.01$), \amar{although the effect is very limited in size and significance}. As shown in Figure~\ref{fig:mbsrq_results}, we observe more pronounced differences in surprise level due to sound condition for people with low MBSRQ scores, who are those less satisfied with their bodies. With regards to strength, people with low MBSRQ scores felt stronger with LF and weaker with HF, while those with high scores felt overall stronger with C than with the altered frequency sounds.  

\textbf{Gait Biomechanics}
We found an interaction with sound condition for number of steps (F=4.04, p=0.019, $\eta_{p}^{2}=0.05$) and mean velocity (F=3.90, p=0.022, $\eta_{p}^{2}=0.05$), \ana{with small effect sizes}. As shown in Fig.~\ref{fig:mbsrq_results}, participants with higher levels of MBSRQ moved faster in HF compared to LF, while participants with low levels of MBSRQ, moved faster in LF compared to HF and C.

\textbf{Lateral Hip and Shoulder Sway}
MBSRQ score had a double interaction with sound condition for hip flexion angle (F=3.02, p=0.049, $\eta_{p}^{2}=0.04$), \amar{however, this effect is small in size}. As shown in Fig.~\ref{fig:mbsrq_results}, higher levels of hip flexion angle are associated with higher values of MBSRQ: LF and C have greater values of hip flexion with high MBSRQ, while for low values of MBSRQ hip flexion values are higher for HF.

\textbf{Electrocardiac Activity}
There was a significant interaction between sound condition and  MBSRQ score for heart rate values (F=4.10, p=0.018, $\eta_{p}^{2}=0.06$), \ana{with a medium effect size}. For high values of MBSRQ, lower values of heart rate were found in HF, while with low levels of MBSRQ the heart rate is \ana{higher for HF}, as shown in Fig.~\ref{fig:mbsrq_results}.

\subsubsection{Differences According to Sensory Imagery Vividness} The interaction of sensory imagery vividness with the effects of sound condition was evaluated, looking both at the general sensory imagery vividness as well as the scores for the different sensory scales which we deemed relevant to our context (visual, auditory, cutaneous (or tactile) and kinaesthetic; see Supplementary Results related to other sensory scales). Imagery vividness scores were used as a covariate in the analyses.

\begin{figure*}[ht]
\includegraphics[width=16.5cm]{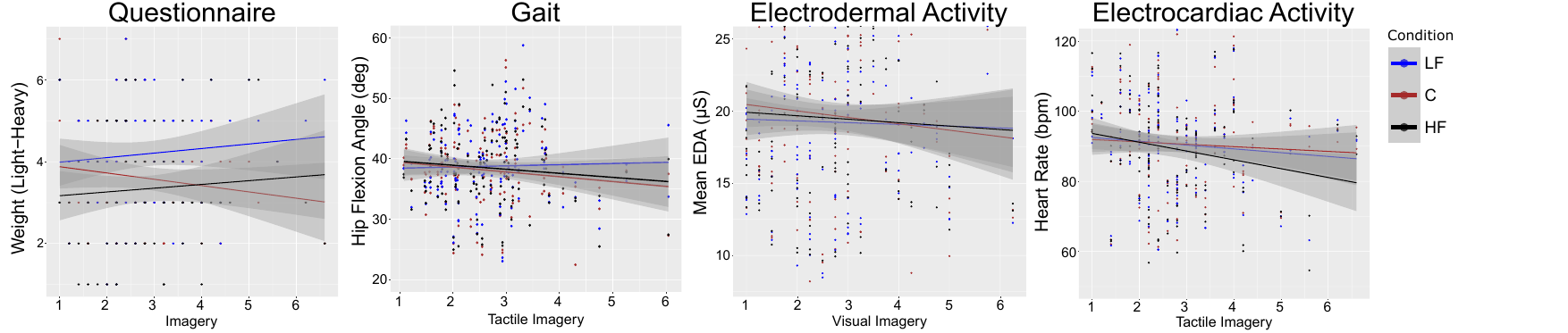}
\caption{Results showing the main differential effects of sound on weight, Hip flexion, electrodermal activity (EDA), electrocardiac activity (ECG) due to differences in sensory imagery (score used as a covariate).}
\label{fig:imagery_results}
\Description{Picture composed of a series of 4 plots showing variation of weight (questionnaire item), Hip flexion, electrodermal activity (EDA),   electrocardiac activity (ECG) for the three sound conditions varying values of imagery vividness}
\end{figure*}

\textbf{Questionnaire Results}
For the question on felt weight, we found an interaction between tactile imagery vividness score, repetition and sound condition (F=3.37 p=0.037, $\eta_{p}^{2}=0.04$), \ana{but the size of the effect was small}. Separate analyses for each repetition on the perceived weight scores showed \amar{small, yet significant,} interaction effects only for the second repetition (F=1.33, p=0.049, $\eta_{p}^{2}=0.01$). As shown in Figure~\ref{fig:imagery_results}, we observe more pronounced differences in felt body weight due to sound condition for people with high tactile imagery scores, who felt the heaviest with LF; in those with low imagery scores we observe they felt the lightest with HF. 

\textbf{Gait Biomechanics}
We found an interaction with sound condition for maximum step acceleration for the auditory imagery score (F=3.33, p=0.038, $\eta_{p}^{2}=0.04$), \ana{with a small effect size}. We found an overall decrease in acceleration for those participants with high values of auditory imagery, in particular for the HF and LF; for those with low auditory imagery score, acceleration was reduced for LF (see Supplementary Results). 

\textbf{Lateral Hip and Shoulder Sway}
There was an interaction between vividness values and sound condition for hip extension angle (F=3.41, p=0.035, $\eta_{p}^{2}=0.05$) \ana{with a small effect size}. As shown in Fig.~\ref{fig:imagery_results}, higher values of tactile imagery produced more pronounced differences in hip sway, with higher values of hip flexion for LF compared to the HF.

\textbf{Electrodermal Activity}
There was a significant interaction between sound condition and visual imagery for mean EDA values (F=3.40, p=0.036, $\eta_{p}^{2}=0.5$), \ana{with a notably large effect size}. Despite lower mean EDA values for elevated visual imagery, both HF and LF provide higher values of EDA compared to C, as shown in Fig.~\ref{fig:imagery_results}.

\textbf{Electrocardiac Activity}
There was a significant interaction between sound condition and tactile imagery for heart rate (F=4.46, p=0.013, $\eta_{p}^{2}=0.06$), \ana{with a medium effect size}, and for the number of successive RR intervals that differ by more than 50ms (F=4.58, p=0.037 $\eta_{p}^{2}=0.50$), \ana{with a notably large effect size}. As shown in Fig.~\ref{fig:imagery_results}, acutely lower values of heart rate were found for participants with high tactile imagery for HF, compared to LF and C.

\section{Discussion}

In this study we explored how SoniWeight Shoes alter body weight perceptions, to better understand what and how individual attributes could be targeted for future technologies enhancing positive body perceptions. Our results confirmed that altered footstep sounds influence participants' perceptions, behavior and movements. Table 3 in Supplementary Material and Table ~\ref{fig:tableResults_main} summarize the effects observed respectively in all the studies with this illusion and in this study, while here we discuss the most relevant results in relation to existing literature.  It should be noted that some of the effects found are relatively small (see Results). In some cases, however, small effect sizes can be important while large ones unimportant~\cite{Morris2020MisunderstandingsAO}, therefore some of these findings should be cautiously considered or ideally reproduced.

We replicated previous findings for the LF and HF conditions. LF sounds contributed to sensations of increased body weight, a more masculine self-perception, and reduced foot acceleration. HF sounds contributed to sensations of reduced body weight, perceived weakness, a heightened sense of femininity, feelings of happiness, increased speed, and a straighter posture. Further, HF feedback resulted in a higher gait speed compared to LF feedback, aligning with the increased feeling of speed observed in the HF condition. Such feedback also prompted greater activation of leg muscles and significant changes in heart activity. Specifically, heart rate decreased and HRV increased in the HF compared to the other conditions. Reduced heart rate has been related to increased attention and relaxation~\cite{Andreassi,Dillon2000}, while increased HRV has been linked to positive emotional states related to safeness and contentment~\cite{DUARTE2017284}. Together, these seem to be physiological pointers of a positive emotional effect of the HF condition. These emotional markers are essential as they have been posited as critical aspects for technology to influence lifestyle changes such as PA~\cite{Biddle2007,Rick2022, Kersten-vanDijk2017}. The differential effects of the LF and HF conditions were confirmed with positional specificity through a novel method involving body maps. More pronounced changes were reported in the lower body for LF, while for HF greater effects were found in the upper body, including the inner arms and shoulders (see Fig. \ref{fig:bmaps_results}).

As hypothesized, this study uncovers a differential effect of altered footstep sounds on physiological, affective, and behavioral levels according to individual factors. The body maps located specific areas with distinct changes for each group. We elaborate on the key interactions identified and their potential implications for the field, and offer insights into design and methodology considerations.

\subsection{Revisiting the Hypotheses}
\textbf{Impact of Eating Disorder Symptomatology (Hypothesis 1)}
Our study revealed that SED influenced how sound feedback impacts body feelings, emotional responses, and gait behavior. Note that our criteria for HIGH and LOW EDEQ are relative to the general population and not a clinical sample (see Supplementary Materials). We specifically found that individuals with HIGH EDEQ, in contrast to those with LOW EDEQ, experienced a decreased ability to localize their feet in both frequency-altered conditions compared to C. Also in those conditions, HIGH EDEQ participants showed an increased EDA, which is indicative of heightened arousal~\cite{boucsein2012electrodermal}. For the same group, the LF condition induced stronger feelings of dominance and a more vivid body experience. These findings are novel, as previous research neither detected significant SED-related effects in self-report measures, potentially due to smaller sample sizes, nor measured EDA~\cite{tajadura-jimenez_body_2022}. 

Separate analyses considering the EDEQ scores as a covariate (see Supplementary Results) showed, similarly to~\cite{tajadura-jimenez_body_2022}, that individuals with higher EDEQ displayed lower acceleration in the HF. This has been linked to a perceived heavier body~\cite{ko2010characteristic,troje2008retrieving}. In contrast, people with lower EDEQ values exhibited a reduced acceleration only in the LF condition aligning with previous studies in healthy participants~\cite{tajadura-jimenez_as_2015}. This, together with the reported loss in feet proprioception and increased arousal in the frequency-altered conditions, point at an influence of sound bodily cues specific to participants with heightened EDEQ levels. A tentative interpretation~\cite{tajadura-jimenez_body_2022} links the unexpected effects in HF to uncertainty intolerance~\cite{Konstantellou2010,Sternheim2011}, a trait often present in people with SED~\cite{Brown2017} characterized by negative beliefs and reactions in response to uncertain circumstances. Such a reaction may push people with SED to feel heavier, a quality often considered negative in SED, even when the HF condition is expected to signal the opposite.
Higher EDEQ values were further associated with an increased lateral hip sway in the HF, and overall reduced leg muscle activation and thorax sway. More research into these motor patterns is necessary to understand the underlying processes and potentially contribute to treatment and diagnostic strategies. 

\textbf{Impact of Level of Physical Activity (Hypothesis 2)}
Our results evidence an influence of PA levels on how sound feedback affects body feelings, emotional responses, and gait behavior. Inactive participants generally experienced more pronounced body feelings, also reporting feeling happier with HF and inversely for LF. Active participants, on the other hand, generally did not exhibit changes in body feelings  between conditions, except for higher levels of reported arousal in the HF condition. Considering the IPAQ score as a covariate (see Supplementary Results), we found that IPAQ values are correlated positively with increased muscle activation, lateral hip sway, and higher HRV. The latter aligns with prior research by~\cite{Quintana2012}. Notably, increases in HRV have been linked to increased fitness levels in middle-aged adults~\cite{NavarroLomas2022,DONG2016}, good cardiovascular health~\cite{Villafaina2021}, and feelings of safeness and content~\cite{DUARTE2017284}. Furthermore, individuals with higher PA levels showed increased leg muscle activations for HF sounds, while for LF sounds they showed a reduced hip sway (indicative of a less stereotypically feminine gait~\cite{Mather1994}) and increased HRV. 

Previous qualitative research employing movement sonification has suggested that individuals with lower (vs. higher) levels of PA manifest a stronger influence of sonification on body and movement perception and emotional states~\cite{ley-flores_soniband_2021}. Here, the acoustic manipulations also exerted distinct effects on inactive and active participants, with the former showing greater changes in body feelings and reported happiness; and the latter displaying stronger changes in physiological responses, movement, and reported arousal. This may suggest a stronger malleability of body perception for inactive participants, potentially due to less attention to ongoing sensory signals due to their lifestyle. Supporting this, physically active participants on the other hand generally showed more pronounced motor and physiological patterns which could indicate a need for more precise monitoring of sensory signals. As suggested by others~\cite{Rick2022, tajadura-jimenez_as_2019, ley-flores_soniband_2021, Singh2017}, the impact on bodily \ana{feelings} and emotional {experience} observed for the inactive group could be potentially targeted to promote engagement in PA and overcome barriers to healthy behaviors. 

\textbf{Impact of Individual Body Concerns (Hypothesis 3)}
Attitudes towards one's body (often referred to as negative/positive body image or body concerns~\cite{Cash1990}) showed an influence on how sound feedback impacts body feelings, gait, and physiological responses. Individuals with higher body concerns (indicated by low scores on the MBSRQ) reported an increase in felt strength in the LF and a decrease in the HF condition, whereas those more satisfied with their bodies reported a reduction in perceived strength with both frequency-altered conditions. Additionally, individuals with higher body concerns reported feeling more surprised with the frequency-altered conditions compared to C. This pattern may be indicative of an increased reliance on prior expectations rather than on ongoing signals~\cite{blanke2012, blanke2015, Apps2014}. Interestingly, for participants with higher body concerns, LF increased mean velocity and HF increased hip sway, while for those with lower body concerns the opposite was observed. Similarly, in the former group, HF led to an increased heart rate, while in the latter, HF resulted in a decreased heart rate. 

While it is early to draw conclusions about underlying mechanisms, our study suggests that body concerns play a pivotal role in shaping the effects of sound on body perception and its associated effects across various dimensions. Considering individual factors related to body concerns (e.g. body appearance, overweight preoccupations, concerns with one's own physical capabilities~\cite{LizanaCaldern2022}), could help understand this intricate relation and develop further strategies.

\textbf{Impact of Gender Perceptions (Hypothesis 4)}
In line with previous findings~\cite{tajadura-jimenez_as_2019}, participants perceived themselves as more feminine and less masculine in the HF condition compared to the LF one. While prior works~\cite{clausen_action_2021} could not confirm a change in walking behavior between conditions, here we found increased hip flexion angles for LF. Furthermore, we found that the overall degree of reported masculinity/femininity after the experimental conditions was correlated with gait patterns related to these gender traits for lateral hip and chest sway~\cite{Mather1994}. 

~\citet{clausen_action_2021} argued that perceived body weight, masculinity/femininity and strength may contribute to the complex construct of gender identity, e.g. lower strength is stereotypically associated with femininity~\cite{Wood2015}. Importantly, this and other associations to body sensations are shaped by sociocultural norms and body ideals~\cite{Thompson2001}, changing across cultures and even groups of people~\cite{TajaduraJimnez2022, Koteles2021}, and should not be considered universal nor static. In fact, our study shows the relatively rapid malleability of gender attributes, contributing to its understanding as a fluid trait (cf.,~\cite{tacikowski_fluidity_2020,Bolt2021, Provenzano2023}). 

\textbf{Impact of Sensory Imagery (Hypothesis 5)}
Adding to the evidence that individual factors influence participants' responses to sound feedback, sensory imagery capacities also had an effect. Worth noting is the relatively stronger influence of tactile imagery (see Results and Supplementary Materials for the effects of other modalities). Participants with high tactile imagery reported feeling particularly heavier during the LF condition. Those with poorer tactile imagery perceived themselves as lighter during the HF feedback, showed an overall decrease in heart rate also during HF, and an increased lateral hip sway during LF. To our knowledge, research is needed on the link between tactile imagery and other aspects of perception (see e.g.,~\cite{ODOWD2022}), and this may be a step in that direction. 

\begin{table*}[ht]
\caption{Results summary for the effects of sound on overall population, individual differences and other correlation results.}

\begin{tabular}{llllllll}
\hline
Measure                                                        & Effect                                              & Overall Population                             & \multicolumn{4}{l}{Individual Differences}                                                                                                                                                                   \\ \hline
\multicolumn{1}{|l|}{}                                         & \multicolumn{1}{l|}{}                               & \multicolumn{1}{l|}{}                          & \multicolumn{1}{l|}{EDEQ}                       & \multicolumn{1}{l|}{IPAQ}                        & \multicolumn{1}{l|}{MBSRQ}                     & \multicolumn{1}{l|}{Imagery}                              \\ \hline
\multicolumn{1}{|l|}{\multirow{12}{*}{Questionnaire}}          & \multicolumn{1}{l|}{Quickness}                      & \multicolumn{1}{l|}{\checkmark} & \multicolumn{1}{l|}{}                          & \multicolumn{1}{l|}{}                          & \multicolumn{1}{l|}{}                          & \multicolumn{1}{l|}{}                                     \\ \cline{2-7} 
\multicolumn{1}{|l|}{}                                         & \multicolumn{1}{l|}{Weight}                         & \multicolumn{1}{l|}{\checkmark} & \multicolumn{1}{l|}{}                          & \multicolumn{1}{l|}{}                          & \multicolumn{1}{l|}{}                          & \multicolumn{1}{l|}{}                                     \\ \cline{2-7} 
\multicolumn{1}{|l|}{}                                         & \multicolumn{1}{l|}{Strength}                       & \multicolumn{1}{l|}{\checkmark} & \multicolumn{1}{l|}{}                          & \multicolumn{1}{l|}{}                          & \multicolumn{1}{l|}{\checkmark} & \multicolumn{1}{l|}{}                                     \\ \cline{2-7} 
\multicolumn{1}{|l|}{}                                         & \multicolumn{1}{l|}{Straightness}                   & \multicolumn{1}{l|}{\checkmark} & \multicolumn{1}{l|}{}                          & \multicolumn{1}{l|}{\checkmark} & \multicolumn{1}{l|}{}                          & \multicolumn{1}{l|}{}                                     \\ \cline{2-7} 
\multicolumn{1}{|l|}{}                                         & \multicolumn{1}{l|}{Masculinity}                    & \multicolumn{1}{l|}{\checkmark} & \multicolumn{1}{l|}{}                          & \multicolumn{1}{l|}{}                          & \multicolumn{1}{l|}{}                          & \multicolumn{1}{l|}{}                                     \\ \cline{2-7} 
\multicolumn{1}{|l|}{}                                         & \multicolumn{1}{l|}{Proprioception}                 & \multicolumn{1}{l|}{\checkmark} & \multicolumn{1}{l|}{\checkmark} & \multicolumn{1}{l|}{}                          & \multicolumn{1}{l|}{}                          & \multicolumn{1}{l|}{}                                     \\ \cline{2-7} 
\multicolumn{1}{|l|}{}                                         & \multicolumn{1}{l|}{Vividness}                      & \multicolumn{1}{l|}{}                          & \multicolumn{1}{l|}{\checkmark} & \multicolumn{1}{l|}{}                          & \multicolumn{1}{l|}{}                          & \multicolumn{1}{l|}{}                                     \\ \cline{2-7} 
\multicolumn{1}{|l|}{}                                         & \multicolumn{1}{l|}{Surprise}                       & \multicolumn{1}{l|}{\checkmark} & \multicolumn{1}{l|}{}                          & \multicolumn{1}{l|}{\checkmark} & \multicolumn{1}{l|}{\checkmark} & \multicolumn{1}{l|}{}                                     \\ \cline{2-7} 
\multicolumn{1}{|l|}{}                                         & \multicolumn{1}{l|}{Agency}                         & \multicolumn{1}{l|}{\checkmark} & \multicolumn{1}{l|}{}                          & \multicolumn{1}{l|}{}                          & \multicolumn{1}{l|}{}                          & \multicolumn{1}{l|}{\checkmark (Tactile)}  \\ \cline{2-7} 
\multicolumn{1}{|l|}{}                                         & \multicolumn{1}{l|}{Valence}                        & \multicolumn{1}{l|}{\checkmark} & \multicolumn{1}{l|}{}                          & \multicolumn{1}{l|}{\checkmark} & \multicolumn{1}{l|}{}                          & \multicolumn{1}{l|}{}                                     \\ \cline{2-7} 
\multicolumn{1}{|l|}{}                                         & \multicolumn{1}{l|}{Arousal}                        & \multicolumn{1}{l|}{}                          & \multicolumn{1}{l|}{}                          & \multicolumn{1}{l|}{\checkmark} & \multicolumn{1}{l|}{}                          & \multicolumn{1}{l|}{}                                     \\ \cline{2-7} 
\multicolumn{1}{|l|}{}                                         & \multicolumn{1}{l|}{Dominance}                      & \multicolumn{1}{l|}{}                          & \multicolumn{1}{l|}{\checkmark} & \multicolumn{1}{l|}{}                          & \multicolumn{1}{l|}{}                          & \multicolumn{1}{l|}{}                                     \\ \hline
\multicolumn{1}{|l|}{Body Visualizer}                          & \multicolumn{1}{l|}{Avatar size (weight parameter)} & \multicolumn{1}{l|}{\checkmark} & \multicolumn{1}{l|}{}                          & \multicolumn{1}{l|}{}                          & \multicolumn{1}{l|}{}                          & \multicolumn{1}{l|}{}                                     \\ \hline
\multicolumn{1}{|l|}{\multirow{5}{*}{Gait}}                    & \multicolumn{1}{l|}{Mean Velocity}                  & \multicolumn{1}{l|}{}                          & \multicolumn{1}{l|}{\checkmark} & \multicolumn{1}{l|}{}                          & \multicolumn{1}{l|}{\checkmark} & \multicolumn{1}{l|}{}                                     \\ \cline{2-7} 
\multicolumn{1}{|l|}{}                                         & \multicolumn{1}{l|}{Mean Acceleration}              & \multicolumn{1}{l|}{}                          & \multicolumn{1}{l|}{\checkmark} & \multicolumn{1}{l|}{}                          & \multicolumn{1}{l|}{}                          & \multicolumn{1}{l|}{}                                     \\ \cline{2-7} 
\multicolumn{1}{|l|}{}                                         & \multicolumn{1}{l|}{Maximum Acceleration}           & \multicolumn{1}{l|}{\checkmark} & \multicolumn{1}{l|}{}                          & \multicolumn{1}{l|}{}                          & \multicolumn{1}{l|}{}                          & \multicolumn{1}{l|}{\checkmark (Auditory)} \\ \cline{2-7} 
\multicolumn{1}{|l|}{}                                         & \multicolumn{1}{l|}{Thorax Extension}               & \multicolumn{1}{l|}{}                          & \multicolumn{1}{l|}{\checkmark} & \multicolumn{1}{l|}{}                          & \multicolumn{1}{l|}{}                          & \multicolumn{1}{l|}{}                                     \\ \cline{2-7} 
\multicolumn{1}{|l|}{}                                         & \multicolumn{1}{l|}{Hips Flexion}                   & \multicolumn{1}{l|}{\checkmark} & \multicolumn{1}{l|}{\checkmark} & \multicolumn{1}{l|}{\checkmark} & \multicolumn{1}{l|}{\checkmark} & \multicolumn{1}{l|}{\checkmark (Global)}   \\ \hline
\multicolumn{1}{|l|}{EMG}                                      & \multicolumn{1}{l|}{Muscle Activation}              & \multicolumn{1}{l|}{\checkmark} & \multicolumn{1}{l|}{\checkmark} & \multicolumn{1}{l|}{\checkmark} & \multicolumn{1}{l|}{}                          & \multicolumn{1}{l|}{}                                     \\ \hline
\multicolumn{1}{|l|}{Electrodermal Activity}                   & \multicolumn{1}{l|}{Mean EDA}                       & \multicolumn{1}{l|}{}                          & \multicolumn{1}{l|}{\checkmark} & \multicolumn{1}{l|}{\checkmark} & \multicolumn{1}{l|}{}                          & \multicolumn{1}{l|}{\checkmark (Visual)}   \\ \hline
\multicolumn{1}{|l|}{\multirow{3}{*}{Electrocardiac Activity}} & \multicolumn{1}{l|}{Heart Rate}                     & \multicolumn{1}{l|}{\checkmark} & \multicolumn{1}{l|}{}                          & \multicolumn{1}{l|}{}                          & \multicolumn{1}{l|}{\checkmark} & \multicolumn{1}{l|}{\checkmark (Tactile)}  \\ \cline{2-7} 
\multicolumn{1}{|l|}{}                                         & \multicolumn{1}{l|}{Heart Rate Standard Deviation}  & \multicolumn{1}{l|}{\checkmark} & \multicolumn{1}{l|}{}                          & \multicolumn{1}{l|}{}                          & \multicolumn{1}{l|}{}                          & \multicolumn{1}{l|}{}                                     \\ \cline{2-7} 
\multicolumn{1}{|l|}{}                                         & \multicolumn{1}{l|}{Heart Rate RMSSD}               & \multicolumn{1}{l|}{\checkmark} & \multicolumn{1}{l|}{}                          & \multicolumn{1}{l|}{\checkmark} & \multicolumn{1}{l|}{}                          & \multicolumn{1}{l|}{}                                     \\ \hline
\end{tabular}
\label{fig:tableResults_main}
\end{table*}

\subsection{Design Takeaways And Inspirations}

\textbf{Concrete design takeaways in basis to the hypotheses' results.} 
Our study shows that the influence of sound feedback to alter body perception seems to be mediated by several individual factors. Most literature on bodily illusions has so far neglected the role of individual differences, which is a limitation when designing devices to be used beyond laboratory settings. Our findings might offer preliminary guidelines or inspirations when designing for specific populations with varying degrees of SED (Hypothesis 1), PA (Hypothesis 2), body concerns (Hypothesis 3) and even imagery capacities (Hypothesis 5). The concrete findings to be used for future designs and research can be found in Table 3 of Supplementary Material, together with other literature and related insights. It should be noted we did not study a clinical sample, and future work is required when engaging with clinical populations. 

While certain effects are consistently observed across various studies and populations (see Table 3 in Supplementary Material), some differ or are exclusively observed according to individual factors. It is crucial to address who may be excluded, unaffected or affected in different ways by these technologies and interventions. We also suggest going beyond the above-mentioned individual factors and also explore associations rooted in physical, sensory and socio-cultural factors to achieve desired effects. By physical we refer to body characteristics changing through ontogeny (e.g., during pregnancy or adolescence, both periods of rapid physical change). By sensory aspects, we refer to bottom-up multisensory associations developed since childhood (e.g., the correlation between sound pitch changes and variations in height and size~\cite{deroy_fernandez_prieto_navarra_spence_2018}, observed in nature~\cite{parise2014}, adult/child and male/female voices~\cite{tajadura-jimenez_embodiment_2017}, musical instrument construction, etc). And socio-cultural top-down processes can be exemplified by the link between high-heels and female stereotypes~\cite{Tonetto2014} and other cultural connotations. Implicitly, these associations are often reinforced through various media forms like games, films and the design of cartoon characters. We suggest to build on these physical, sensory, and socio-cultural associations, and to design flexible technologies that are customizable, adaptable, and programmable. 

\textbf{Technical design takeaways.} In contrast to the analog system employed in previous footsteps illusion experiments~\cite{tajadura-jimenez_as_2015, tajadura-jimenez_as_2019, tajadura-jimenez_bodily_2017, TajaduraJimnez2022, clausen_action_2021, brianza_as_2019}, SoniWeight Shoes are wireless and lighter. Such features, together with its relatively affordable cost and optimal quality with no perceivable latency (1.6ms), are important to widen its applicability and mobility. Based on the original analog device~\cite{tajadura-jimenez_as_2015}, a careful configuration of the digital filters was carried out to reproduce the LF and HF conditions, yielding a set of numerical z-transform H(z) coefficients that allow their replicability across digital devices. This version makes programming the digital processing accessible, enabling, e.g., a dynamic or gradual application of the filters (HF or LF), or easily personalizing the cutoff frequency according to individual needs. \laia{Others can derive technical inspiration from this implementation to design ubiquitous systems.}

\subsection{Methodological Reflections, Limitations and Future Work} 
Our use of body maps addresses bodily changes with spatial accuracy. The intricate responses emphasize the need for tailored approaches in technology design and the potential of such a method to address, delineate and eventually target specific body areas. This could be a methodological contribution to future work in HCI and potentially be linked to other markers (e.g. physiological or behavioral), for example using machine learning to address more complex interrelations. Complementary qualitative approaches could contribute to understanding the intricate nature of this phenomenon. For example, in-depth interviews or focus group discussions, could stem a richer narrative of participants' feelings, thoughts, and experiences~\cite{anne_cochrane_body_2022,turmo_vidal_towards_2023}. Further, marking different levels of intensity of felt changes (through e.g. different colours) could enrich the findings. \enlargethispage{10pt}

Worth noting from our study is that while wearing the Rokoko suit might have influenced participants' sensations, unlike studies aiming at somatosensory stimulation and body image treatment with tight neoprene (e.g.,~\cite{Grunwald2005}), here the suit was loosely worn. In any case, our within-subjects design should ensure that any effects are due to our manipulated variables. As a related note, the numerous measures applied in our study could seem like a cognitive burden to participants. However, the majority of (longer) questionnaires were administered beforehand and the sensor data was collected implicitly and automatically. Only the avatar task, questionnaire on body feelings, and body map were applied after each walking trial, a comparable amount to other studies in the field (e.g.,~\cite{tajadura-jimenez_as_2015,botvinick_rubber_1998, roel_lesur_different_2023}).

Technically, several improvements to SoniWeight Shoes are envisioned. First, the addition of binaural immersive audio. While currently the signal captured at the feet is emitted directly to the ears (through headphones); in a real situation, our auditory perspective and the distance to our feet would be accounted for. This can be simulated using HRTFs~\cite{Rafaely2022} personalized for each user~\cite{Zhou2021} based on their anthropometric features and applied through digital processing. Open headphones, bone conducting~\cite{Valjamaebone}, or earbuds could be integrated to avoid isolation from the acoustic environment. Further, the integration of wireless devices (microphones and headphones) would improve comfort and portability. Lastly, applying a real-time noise canceling algorithm to filter out spectral content external to the footsteps would be an important step for its use in real-life scenarios. 

Previous research on the footsteps illusion~\cite{tajadura-jimenez_as_2019} revealed that changes in body perceptions tend to dissipate once the feedback ceases, a characteristic in line with the rapid adaptability of body perceptions~\cite{botvinick_rubber_1998}; however, that study showed that body feelings or behavioral alterations exhibited longer-lasting effects. Relatedly, research on other sensory-driven illusions has demonstrated enduring psychological impacts following brief exposures, such as changes in attitudes towards others~\cite{maister_changing_2015}. Hence, despite the brevity of exposure, it may still influence subsequent body-related feelings and behaviors. This holds promise for addressing negative body perceptions and overcoming barriers to healthy behaviors~\cite{Rick2022} by e.g., reshaping attitudes and emotions related to one's body~\cite{ley-flores_soniband_2021,Singh2017}. However, longer-term and in-the-wild studies are necessary as body perceptions are highly context-dependent. Promising examples of such approaches are utilizing sonification technologies at home for extended periods (2-4 weeks) in the context of PA~\cite{ley-flores_soniband_2021} or physical rehabilitation~\cite{Singh2017}. Nevertheless, these studies predominantly employed qualitative methods, leaving the quantitative evaluation of changes in behavior, physical/physiological responses, or emotional states yet to be explored. Any potential adaptability after prolonged exposure remains to be studied. 

Based on the prevalence and impact of eating disorders and physical inactivity, we aimed to study how wearable acoustic interventions can be targeted for these groups. We have thus extended our sample compared to similar studies, however, it is worth recognizing that it was primarily composed of university students. While our results are an important step for addressing individual differences, we worked with a non-clinical sample and a limited population group. Clinical populations may show even more distinct patterns than our sample and eventually benefit from these approaches, but future research is necessary. Studying a more diverse population involving other individual factors might be deemed important both clinically and in understanding the mechanisms behind population differences in body perception. 

\subsection{Novelty and Contributions}
This paper broadens the scope in prior HCI research on sound feedback to alter body perception~\cite{tajadura-jimenez_as_2015,tajadura-jimenez_embodiment_2017,tajadura-jimenez_as_2019,tajadura-jimenez_body_2022,brianza_as_2019,gomez-andres_enriching_2020,clausen_action_2021}. Beyond the contributing with quantitative body maps for HCI ~\cite{anne_cochrane_body_2022, turmo_vidal_towards_2023}, we highlight three main contributions. 

First, we provide evidence that varying degrees of SED, PA, and sensory imagery contribute to the effects of footstep sonification. This adds to the literature on individual factors on body perception, including SED~\cite{Eshkevari2012, Keizer2014, tajadura-jimenez_body_2022},  PA~\cite{ley-flores_soniband_2021}, and gender perceptions and aspirations ~\cite{tajadura-jimenez_as_2019, clausen_action_2021}. Based on this, we outlined design opportunities considering both bottom-up multisensory processes and top-down prior knowledge and beliefs~\cite{botvinick_rubber_1998, blanke2012, blanke2015, Apps2014}.

Second, using an improved sonification device, we replicated previous findings both for the general population~\cite{tajadura-jimenez_as_2015} and people with relatively high levels of SED~\cite{tajadura-jimenez_body_2022}. SoniWeight Shoes are a more portable and lighter digital version of the system described in~\cite{tajadura-jimenez_as_2015}. The device does not require AC power, employs carefully evaluated filters conforming to the original analog equalizer, and provides very low latency. With these optimizations the wearable device opens its scope for ubiquitous real-world scenarios.

Third, a dataset involving 84 participants with varying degrees of the individual factors considered, with measures from a variety of tasks and sensors. This might support future analyses, comparisons with other studies, or reviews, and is an important contribution due to the relatively diverse sample in terms of the above criteria.

At large, this work may be relevant for HCI and cognitive science research on body perception, multisensory stimulation, or sonification research. It might support and inspire the design of personalized technologies in rehabilitation, sports, virtual reality/gaming, and beyond. 

\subsubsection{A Call Towards Personalization of Sensory Feedback}
We have highlighted the inherent variability in individual responses to auditory feedback. These findings add to, and strengthen, existing research showing that the malleability of body perception in response to sensory signals is influenced by individual differences. For example, visuo-tactile illusions are experienced differently among younger and older individuals~\cite{TAJADURAJIMENEZLongo2012,Marotta2018,Cowie2013,Cowie2018TheDO,Nava2018, roel_lesur_how_2020}, and those with higher interoceptive sensitivity exhibit less body malleability (\cite{Tsakiris2011, tajadura-jimenez_balancing_2014, DURLIK201442} but see~\cite{ROELLESUR2020}). These individual differences can manifest in various ways, such as sensorial variations (e.g., individuals with SED integrate sensorial bodily signals differently from those without~\cite{tajadura-jimenez_body_2022}, diverse sensory imagery abilities~\cite{MILLER2013140}), emotional differences (e.g., varying body concerns, fear of fragile bones~\cite{Rick2022}), socio-cultural influences (e.g., gender stereotypes, weight stigma, rooted in “ideal body images”~\cite{Thompson2001}), body types~\cite{spiel_bodies_2021}, preferences, and more. 

We here advocate for the integration of individual differences in both research and technology design, emphasizing the need for personalized sensory technologies. While our call serves to highlight the importance of tailoring feedback across populations, future work including design guidelines and considerations would be beneficial. For this, design-oriented works may provide inspiration, through e.g., open-ended sensory feedback technologies that enable multiple sensemakings~\cite{turmo_vidal_intercorporeal_2023, turmo_bodylights} or employing machine learning approaches to cater to fluctuating needs ~\cite{duval_machlearn}. We hope to encourage other researchers to explore the role of individual differences in their own work.

\section{Conclusion}
Our study shows that SoniWeight Shoes, a prototype designed to change footstep sounds with a strong emphasis on portability, convenience, and minimal latency, indeed alters body-weight perception. Variations in individuals' SED, PA and body concerns, all of which are relevant health markers~\cite{Cash1997, Schwarzer1992, Cairney2007, McAuley1993}, influence the effect of sound feedback on body perception, posture, gait, emotional \ana{experience} and physiological markers. Besides a call to consider diverse populations and individual factors and contexts in related research and design, we point at a relevant opportunity for improving well-being through targeting body perceptions. For taking this step into an actionable tool, our work should be extended to clinical samples and longer-term studies in real-world scenarios. We have contributed with a public dataset to serve future research in this direction that can be studied within and beyond HCI (e.g., in cognitive and sport sciences). As a further methodological contribution within HCI, we report a quantitative use of body maps to assess specific bodily regions where salient perceptual changes are reported. This work can contribute to understand and design tailored sensory feedback to alter perceptions, attitudes, emotions, and behaviors related to one's body.

\begin{acks}
This research was supported by the European Research Council (ERC) under the European Union’s Horizon 2020 research and innovation programme (grant agreement No 101002711; project BODYinTRANSIT) and the Spanish Agencia Estatal de Investigación (PID2019-105579RB-I00/AEI/10.13039/5011 00011033; project MAGIC outFIT).  A.V. is a policy analyst in the European Parliamentary Research Service (EPRS), he is writing in a personal capacity, and any views expressed do not represent an official position of the European Parliament.
\end{acks}



\balance
\bibliographystyle{ACM-Reference-Format}
\bibliography{REFS}

\end{document}